\documentclass{article}
\usepackage{hchang}
\usepackage{hyperref}
\usepackage{comment}
\usepackage{tcolorbox}
\usepackage{color}
\usepackage{graphicx} % allow embedded images
  \setkeys{Gin}{width=\linewidth,totalheight=\textheight,keepaspectratio}
  \graphicspath{{graphics/}} % set of paths to search for images
\usepackage{booktabs}
\usepackage{array} % book-quality tables
\usepackage{units}    % non-stacked fractions and better unit spacing
\usepackage{multicol} % multiple column layout facilities
\usepackage{lipsum}   % filler text
\usepackage{fancyvrb} % extended verbatim environments
  \fvset{fontsize=\normalsize}% default font size for fancy-verbatim environments
\usepackage[capitalise]{cleveref}
\usepackage{rotating}
\usepackage{transparent}
\usepackage{multicol}
\usepackage{tikz}
\usetikzlibrary{calc,decorations.pathreplacing,shapes.geometric}
\usepackage{fullpage}
\usepackage{soul}
\usepackage{thmtools}
\usepackage{thm-restate}
\usepackage{algorithm}
\usepackage{algorithmicx}
\usepackage{algpseudocode}
\usepackage{arydshln}


\newtheorem{lemma}{Lemma}
\newtheorem{theorem}{Theorem}

\newtheorem{observation}{Observation}

\newtheorem{remark}{Remark}
\newtheorem{question}{Question}

\newcommand{\draw}{\in_{R}}

\newcommand{\Cc}{\mathcal C}
\newcommand{\cX}{\mathcal{X}}

\newcommand{\real}{\mathbb{R}}
\newcommand{\opt}{\mathsf{Opt}}

\newcommand{\Tsssp}{T_{\mathrm{SSSP}}}

\newcommand{\Reg}{\operatorname{Regret}}

\newcommand{\defn}[1]{\EMPH{#1}}

\newcommand{\cP}{\mathcal{P}}
\newcommand{\cT}{\mathcal{T}}

\newcommand{\bbE}{\mathbb{E}}
\newcommand{\cK}{\mathcal{K}}

\newcommand{\bdry}{\partial}

\newcommand{\lightgray}[1]{\textcolor{gray}{#1}}

\crefname{problem}{problem}{problems}
\Crefname{problem}{Problem}{Problems}
\crefformat{problem}{problem~(#2#1#3)}
\Crefformat{problem}{Problem~(#2#1#3)}

\renewcommand{\leq}{\leqslant}
\renewcommand{\le}{\leqslant}
\renewcommand{\geq}{\geqslant}

\algtext*{EndIf}
\algtext*{EndFor}
\algtext*{EndWhile}
\algtext*{EndProcedure}

\title{Flow-Optimal Separators for Minor-Free and Induced-Minor-Free Graphs
in Linear Time via Follow the Regularized Leader}

\author{
Hung Le\thanks{University of Massachusetts Amherst.  \texttt{hungle@cs.umass.edu}. Supported by the NSF CAREER award CCF-2237288 and the NSF grant CCF-2517033.}
\and
Jack Spalding-Jamieson\thanks{Independent. \texttt{jacksj@uwaterloo.ca}}
}
\date{}

\begin{document}

\maketitle

\begin{abstract}

We give a new algorithm for finding balanced separators for minor-free and induced-minor-free graphs. A key new technical idea is a novel connection to online learning: we design a continuous (non-convex) optimization problem and solve it using the Follow the Regularized Leader (FTRL) framework. Our implementation of each FTRL step yields a purely combinatorial algorithm which simply assigns and adjusts vertex weights over multiple rounds. As a result, we obtain:
\begin{enumerate}
    \item a balanced separator of an $H$-induced-minor-free graph of size $O(\log(|V(H)|)\sqrt{|V(H)| + |E(H)|}\sqrt{m})$ in \EMPH{expected linear time} for a fixed $H$. As a corollary, we obtain an algorithm to find a balanced separator for string graphs of size $O(\sqrt{m})$ in expected linear time without requiring the string representations of the graphs. 
    \item a balanced separator of size $O(h\log h \sqrt{n})$ for $K_h$-minor-free graphs in \EMPH{expected linear time} for a fixed $h$. Our algorithm unifies two long lines of work on separators for minor-free graphs since Alon, Seymour, and Thomas~\cite{AST90}: one optimizes for the separator's size, and the other optimizes the running time.
        It also gives a natural interpretation of the vertex weighting scheme in a recent linear-time algorithm by Bonnet, Korhonen, Le, Li, and Masa{\v{r}}{\'\i}k~\cite{bonnet2026separator}.
\end{enumerate}

The separator bounds for both classes of graphs are flow-optimal: techniques that rely on flow-cut duality, including ours, cannot give balanced separators of (asymptotically) smaller size.

\paragraph{AI Disclosure.}
The key high-level ideas, especially the online learning perspective, of this work were human-devised,
but many of them were fleshed out and developed using LLM assistance.
\EMPH{GPT 5.5} and \EMPH{GPT 5.6-Sol} played useful roles in our initial developments for the presentation using FTRL,
while \EMPH{GPT 6-Astra} improved our initial running time in~\Cref{sec:cut-or-cluster}. Other models, namely \EMPH{GPT 6.1-Sol} and \EMPH{Claude Opus 5.5} have been used to proofread.
The presentation and the write-up are by the human authors only. Of course, we take full responsibility for our results.

\end{abstract}

\section{Introduction}

Separator theorems are cornerstones of modern algorithmic graph theory. They facilitate divide-and-conquer algorithms on graphs, a basic technique in algorithm design. Furthermore, research in separator theorems often reveals deep and unexpected connections between different areas of math and computer science, some of which we will mention here.

\paragraph{Balanced separators of sparse graphs.~} In 1979, Lipton and Tarjan~\cite{LT79}  constructed a balanced separator of size $O(\sqrt{n})$ for $n$-vertex planar graphs, and showed various algorithmic applications of their result~\cite{LT80}. Their main technique is to use the duality of the (planar) embedding, which was also used in a subsequent (and more general) balanced separator result for surface-embedded graphs~\cite{GHT84}. A decade later, Alon, Seymour and Thomas~\cite{AST90} showed how to get around using (planar or surface) embeddings and constructed a balanced separator for any $K_h$-minor-free graph of size $O(h^{3/2}\sqrt{n})$ in time $O(m\sqrt{hn})$. Both the (dependence on $h$ of the) separator size and the running time are not optimal, and have hence motivated two long lines of work that remain active to this day.

One line of work seeks to reduce the separator size to $O(h\sqrt{n})$, while treating the running time as a secondary concern.  Kawarabayashi and Reed~\cite{KR10} used Robertson-Seymour decomposition and claimed to obtain a separator of size  $O(h\sqrt{n} + f(h))$ for some function $f$ of $h$. At around the same time, Biswal, Lee and Rao~\cite{BLR10} devised a new technique for constructing balanced separators via a connection to flow-cut duality. Though their separator's size is worse than that of Alon, Seymour and Thomas~\cite{AST90}, subsequent results~\cite{SpJ25,le2026separator} followed this technique, along with recent advances with padded decomposition for $K_h$-minor-free graphs by Conroy and Filtser~\cite{CF25}, to bring the separator size to a \EMPH{nearly-optimal bound} of $O(h\log h \sqrt{n})$.
This bound is optimal for a class of techniques relying on flow-cut duality and hence $O(h\log h \sqrt{n})$ becomes a barrier.
Recently,
Le~\cite{le2026separator} broke this barrier by combining two combinatorial approaches of Alon, Seymour and Thomas~\cite{AST90} and Plotkin, Rao, and Smith~\cite{PRS94}.
During the final stages of our own work on this paper,
this line of work was fully resolved by a superset of the authors on this paper, with AI assistance,
by building on Le's techniques~\cite{conroy2026khminorfreegraphsbalancedseparators}.
However, a common theme in this entire line of work is that the running time does not improve on that of Alon, Seymour and Thomas~\cite{AST90}.
It is possible that heavy machinery from recent dynamic graph algorithms could improve the running time, though it is unclear what new insights one could gain. Often, a lack of good running-time bounds points to our lack of structural understanding of minor-free graphs.

A second line of work focuses mostly on the regime of $h = O(1)$ and aims to improve the running time to linear, often at the cost of a larger separator, both in terms of $h$ and $n$. Reed and Wood~\cite{RW09} gave the first linear-time algorithm for finding a balanced separator of size $O_h(n^{2/3})$. However, the dependence on $n$ is not optimal. And the dependence on $h$ in both the running time and the separator's size is $2^{O(h^2)}$, so even a very small value of $h$, say $h = \log(n)$, makes both extremely large. For linear-time algorithms, this algorithm remained the state of the art until very recently; Bonnet, Korhonen, Le, Li, and Masa{\v{r}}{\'\i}k~\cite{bonnet2026separator} gave an $O(\poly(h)\cdot n)$ time algorithm that gives a separator of size $O(\poly(h)\cdot\sqrt{n})$.
Their algorithm achieves optimal dependence on $n$ while losing only a $\poly(h)$ factor instead of an exponential factor in $h$. Notably, they introduce a novel technique for constructing balanced separators: it assigns weights to the graph's vertices, and at each step, if the algorithm fails to output a balanced separator, it adjusts the vertex weights for the next step. On the other hand, the $\poly(h)$ factors in both the running time and the separator's size are quite large: $O(h^{13})$.

The two aforementioned lines of work, while productive in pursuing their own goals, diverge in their techniques and bounds: either the running time is too big (even for a fixed $h$) or the separator size is too big. Achieving the best of both is left as an open problem by Bonnet et al.~\cite{bonnet2026separator}:

\begin{question}\label{quest1} Can we construct a balanced separator of (nearly) optimal size in time $O(\poly(h)n)$ for $K_h$-minor-free graphs?
\end{question}

At the conceptual level, \Cref{quest1} asks for a unification of the two long lines of work on separators for $K_h$-minor-free graphs over the last few decades.

Beyond minor-free graphs, Plotkin, Rao, and Smith~\cite{PRS94} introduced the notion of shallow minor-free graphs. Roughly speaking, $K_h$ is an $\ell$-shallow minor of $G$ if the minor model of $K_h$ in $G$ can be formed from (disjoint) subgraphs of radius at most $\ell$.   Plotkin, Rao, and Smith~\cite{PRS94} adapted the separator theorem of Alon, Seymour, Thomas~\cite{AST90} to show that $\ell$-shallow $K_h$-minor-free graphs have balanced separators of size $O(n/\ell + \ell h^2 \log n)$. The concept of shallow minors and the  Plotkin-Rao-Smith  separator theorem are the basis of the (award-winning) theory of graph sparsity by Ne{\v{s}}et{\v{r}}il and Ossona de Mendez~\cite{NOdM12}. One key result, among others, in this theory is a precise characterization of the classes of graphs that have strongly sublinear separators of the form $O(n^{1-\alpha})$ for a constant $\alpha\in (0,1)$~\cite{NOdM12,DN16,ER18}: these are exactly graphs with polynomial expansion\footnote{We refrain from defining this terminology, as it requires non-trivial effort.}.

\paragraph{Balanced separators for dense graphs.} The graph classes mentioned so far are sparse: for instance, every $n$-vertex $K_h$-minor-free graph has $O(n)$ edges, where big-O hides the dependency on the class parameter. Many structural graph classes of practical interest are dense. For example, unit disk graphs, which are the intersection graphs of unit disks on the plane, can contain a large clique.
This example is generalized by the class of all geometric intersection graphs in the plane,
also known as string graphs.
Despite their density, string graphs are $K^{(1)}_5$-induced-minor-free~\cite{ABPT25}, where $K^{(1)}_5$ is the graph obtained from $K_5$ by subdividing each edge once. Here, a graph $H$ is an \EMPH{induced minor} of a graph $G$ if $H$ can be obtained from $G$ by a sequence of vertex deletions and edge contractions\footnote{If we also allowed edge deletions in the sequence, then $H$ would be a minor of $G$.}.
In a remarkable recent work, Korhonen and Lokshtanov~\cite{KL24} gave a separator theorem for all induced-minor-free graphs.

\begin{theorem}[Korhonen and Lokshtanov~\cite{KL24}]\label{thm:KL-separator} Any $H$-induced-minor-free graph $G$ with $n$ vertices and $m$ edges has a balanced separator of size $O(\log n \sqrt{|V(H)| + |E(H)|} \sqrt{m})$ that can be constructed in polynomial time.
\end{theorem}

The $\sqrt{m}$ factor is best possible: $K_n$ has $m = \Theta(n^2)$ edges, and every balanced separator of $K_n$ has size $\Omega(n) = \Omega(\sqrt{m})$.
Their proof introduced a novel flow-based technique that also yielded the nearly optimal bound for balanced separators of minor-free graphs via the flow-based techniques mentioned above~\cite{le2026separator}.  In the same paper, Korhonen and Lokshtanov~\cite{KL24} gave many algorithmic applications of \Cref{thm:KL-separator}; see Corollary 1.2 of~\cite{KL24}.
In these applications, computing the balanced separator is the first step. However, though polynomial, the running time is far from linear because it requires computing a concurrent flow between $\Omega(n^2)$ pairs of vertices.
A recent work~\cite{DRSW27} improved the separator size by Korhonen and Lokshtanov~\cite{KL24} to $O(\log(|V(H)| + |E(H)|) \sqrt{|V(H)| + |E(H)|} \sqrt{m})$ which is also nearly optimal. The key new idea in this work~\cite{DRSW27} is an embedding of vertex-weighted induced-minor-free graphs into the line with a small average distortion.
In other words, a direct (vertex) flow-cut gap bound.
Such an embedding can be used to replace a padded decomposition subroutine in the work of Korhonen and Lokshtanov~\cite{KL24}.
The techniques of both works run in polynomial time (for a relatively large polynomial),
so the overall question of fast running time for such results remains:

\begin{question}\label{question2} Can we construct a balanced separator of (nearly) optimal size $O(\log |V(H)| \sqrt{|V(H)| + |E(H)|} \sqrt{m})$  in time $O(\poly(|V(H)|)(m+n))$?
\end{question}

\subsection{Our results}\label{subsec:results}

Our main results affirmatively answer both \cref{quest1} and \cref{question2}:
we give linear-time algorithms for finding nearly-optimal balanced separators
of both
$K_h$-minor-free graphs (for a fixed $h$)
and
$H$-induced-minor-free graphs (for a fixed graph $H$).

\begin{restatable}{theorem}{minorseparator}\label{thm:minor-separator}
Let $G$ be $K_h$-minor-free with $n$ vertices and $m$ edges.
There is a randomized algorithm that returns a $3/4$-balanced separator
of size $O(h \log h \sqrt{n})$ in expected $O\left(h^7(m+n)\right)$ time.
\end{restatable}
\begin{restatable}{theorem}{indminorseparator}\label{thm:indminor-separator}
Let $G$ be $H$-induced-minor-free with $n$ vertices and $m$ edges.
There is a randomized algorithm that returns a $3/4$-balanced separator
of size $O(\log |V(H)|\sqrt{m(|V(H)| + |E(H)|)})$ in expected
$O\left(|V(H)|^7(m+n)\right)$ time.
\end{restatable}

Our \Cref{thm:minor-separator} not only answers \Cref{quest1} but also unifies the two lines of work on balanced separators for minor-free graphs since Alon, Seymour, and Thomas~\cite{AST90}. Each of our separator sizes are only $\log h$ away from the optimal bound, and both are optimal for techniques that rely on flow-cut duality, including ours.
In particular, both are tight for $3$-regular expander graphs, which are $K_{\Omega(\sqrt{n})}$-minor-free. \Cref{tab:separator-results} compares our results with previous work.

\begin{table}[htbp]
\centering
\small
\setlength{\tabcolsep}{5pt}
\renewcommand{\arraystretch}{1.2}

{%
\setlength{\arrayrulewidth}{1pt}
\setlength{\dashlinedash}{4pt}
\setlength{\dashlinegap}{3pt}
\begin{tabular}{@{}c:c@{}}
\begin{minipage}[t]{0.48\textwidth}
\vspace{0pt}
\centering
\textbf{Sparse Graphs}\par
\medskip
\resizebox{\linewidth}{!}{%
\begin{tabular}{@{}clll@{}}
\toprule
Year & Reference & Separator size & Running time \\
\midrule
\multicolumn{4}{c}{$K_h$-minor-free graphs} \\
\midrule
1990 & \cite{AST90}
 & $O(h^{3/2}\sqrt n)$ & $O(m\sqrt{hn})$ \\
1994 & \cite{PRS94}
 & $O(h\sqrt{n\log n})$ & $O(hm\sqrt{n\log n})$ \\
2008 & \cite{BLR10}
 & $O(h^3\sqrt{n\log h})$ & $\poly(n)$ \\
2009 & \cite{RW09}
 & $2^{O(h^2)}n^{2/3}$ & $2^{O(h^2)}n$ \\
2010 & \cite{KR10}$^\ast$
 & $O(h\sqrt n+f(h))$ & $O(g(h)n^{1+\varepsilon})$ \\
2011 & \cite{WulffNilsen11}
 & $O(\poly(h)n^{4/5+\varepsilon})$ & $O(\poly(h)n)$ \\
2011 & \cite{WulffNilsen11}
 & $O(h\sqrt{n\log n})$ & $O(\poly(h)n^{5/4+\varepsilon})$ \\
2025 & \cite{SpJ25}$^\dagger$
 & $O(h\log h\log\log h\sqrt n)$ & $\poly(n)$ \\
2026 & \cite{bonnet2026separator}
 & $O(h^{13}\sqrt n)$ & $O(h^{13}n)$ \\
2026 & \cite{le2026separator}
 & $O(h\log h\sqrt n)$ & $\poly(n)$ \\
2026 & \cite{le2026separator}
 & $O(h\sqrt{n\log h})$ & $O(h\sqrt{\log h}\,n^2+h^4m)$ \\
2026 & \cite{kolmogorov2026fastapproximationalgorithmminimum}
 & $O(h\log h\sqrt n)$ & $O((m+n)^{1+\varepsilon})$ \\
2026 & \cite{conroy2026khminorfreegraphsbalancedseparators}
 & $O(h\sqrt n)$ & Not stated \\
\multicolumn{2}{c}{\textbf{This paper}}
 & $O(h\log h\sqrt n)$ & $O(h^7(m+n))$ \\
\bottomrule
\end{tabular}%
}
\end{minipage}
&
\begin{minipage}[t]{0.48\textwidth}
\vspace{0pt}
\centering
\textbf{Dense Graphs}\par
\medskip
\resizebox{\linewidth}{!}{%
\begin{tabular}{@{}clll@{}}
\toprule
Year & Reference & Separator size & Running time \\
\midrule
\multicolumn{4}{c}{String graphs} \\
\midrule
2010 & \cite{fox2010separator}
 & $O(m^{3/4}\sqrt{\log m})$ & Not stated \\
2014 & \cite{matouvsek2014near}
 & $O(\sqrt m\log m)$ & $\poly(n)$ \\
2017 & \cite{lee:LIPIcs.ITCS.2017.1}
 & $O(\sqrt m\log m)$ & $\poly(n)$ \\
2026 & \cite{DRSW27}
 & $O(\sqrt m)$ & $\poly(n)$ \\
2026 & \cite{kolmogorov2026fastapproximationalgorithmminimum}
 & $O(\sqrt m)$ & $O((m+n)^{1+\varepsilon})$ \\
\multicolumn{2}{c}{\textbf{This paper}}
 & $O(\sqrt m)$ & $O(m+n)$ \\
\midrule
\multicolumn{4}{c}{$H$-induced-minor-free graphs, $h_V=|V(H)|$, $h_E=|E(H)|$} \\
\midrule
2024 & \cite{KL24}
 & $O(\sqrt{(h_V+h_E)m}\log n)$ & $\poly(n)$ \\
2026 & \cite{DRSW27}
 & $O(\sqrt{(h_V+h_E)m}\log h_V)$ & $\poly(n)$ \\
\multicolumn{2}{c}{\textbf{This paper}}
 & $O(\sqrt{(h_V+h_E)m}\log h_V)$ & $O(h_V^7(m+n))$ \\
\bottomrule
\end{tabular}%
}
\end{minipage}
\end{tabular}
}

\caption{Balanced separator bounds for graphs with $n$ vertices and $m$ edges.
Here $\varepsilon>0$ is an arbitrary fixed constant,
and $f(h),g(h)$ are fast-growing functions from the graph minor structure theorem.
Our running times are in expectation.
$^\ast$Full proof unavailable (see \cite{bonnet2026separator,le2026separator} for more nuanced discussions).
$^\dagger$A corollary, distinct from the main algorithm the paper studies.}
\label{tab:separator-results}
\end{table}

Both of our results are based on the same set of underlying ideas.
At a very high level, our approach may seem quite standard:
We devise a continuous optimization problem, which we solve using an online learning algorithm.
In the course of executing this online learning algorithm,
we may terminate early to output a separator.
If we never terminate early, we instead certify the existence of a minor or induced-minor.
This kind of approach is quite similar to many existing approximation algorithms
for graph partitioning (e.g.~\cite{arora2016combinatorial,lau2024fast,kolmogorov2026fastapproximationalgorithmminimum}).
However, there are two key differences that prevent us from using
much of the existing machinery:
The continuous optimization problem we devise is non-convex,
and the online learning algorithm we employ has not yet seen
any use in combinatorial approximation algorithms or graph partitioning (to our knowledge).

We discuss the online learning algorithm first.
To design our algorithms,
we introduce a new approach based on the follow the regularized leader (FTRL) framework, a well-studied technique in online learning and convex optimization.
For readers who are not familiar with online learning and FTRL, we give a very short primer in \Cref{subsec:online-FTRL} by outlining basic ideas in the FTRL framework.
A very short summary is: learning happens in $T$ rounds, where in each round $t$, a learner has to output a vector $x_t \in \cX\subseteq \real^d$, and suffers a loss $\ell_t(x_t)$ for a loss function $\ell_t: \cX\rightarrow (-\infty,+\infty]$ that the learner does not know. ($\cX$ is the domain.)
In future rounds, the learner gets access not just to the previous losses, but also their gradients.
The learner's goal is to minimize regret, defined as the difference between the learner’s cumulative loss and the cumulative loss of the best fixed vector in hindsight.
The FTRL framework would tell the learner how to choose $x_t$ based on the previous choices $x_1,\ldots, x_{t-1}$ and the loss functions at previous steps $\ell_1(x),\ldots, \ell_{t-1}(x)$. In online learning, the loss functions are chosen by an online adversary.
FTRL itself is a generalization of many popular online learning algorithms,
including multiplicative weight update and matrix multiplicative weight update.
However, the specific instantiation of FTRL that we will employ
is not captured by these special cases.

Essentially any constrained convex optimization problem can be set up as a two-player
min-max game via Lagrangian duality, so online learning is also frequently employed as a method for solving convex optimization problems. Since solutions to convex relaxations of combinatorial problems can often be rounded
to approximate solutions of the original combinatorial problems,
one may expect that this is the methodology with which
online learning is employed for approximation algorithms.
However, doing things in this manner does not make use of the special characteristics of online
learning.
Instead, many fast approximation algorithms
use the idea of \emph{early termination}:
Perform a learning game over the course of (at most) $T$ rounds
between a learner and an adversary,
where the adversary has the option of terminating the entire game prematurely
and outputting a desired combinatorial solution
(see e.g.~\cite{arora2016combinatorial,lau2024fast,kolmogorov2026fastapproximationalgorithmminimum}).
In such games (including all the examples),
the adversary will also often use loss functions with some helpful properties
like sparse gradients,
in order to reduce the computational overhead of the learner.
This is often done in order to reduce the time complexity of the learner
from a large polynomial to near-linear (per-round).
If the learning game completes all $T$ rounds,
such approaches then
certify a lower bound on the value of the problem.
These lower bounds are guaranteed via duality and standard regret guarantees.
Our algorithm shares a number of similar ideas with such works.
However,
unlike the cited works,
our objective will not simply follow as a natural convex relaxation of a combinatorial problem.
In fact, as previously mentioned, it will not be convex at all,
which means
duality and standard regret guarantees will not help us in proving our results.

At a high level, our goal, which is also our major conceptual contribution, is to set up an online learning game with early termination such that, for a suitable choice of $T$ and the loss functions,
learning over $T$ rounds yields an early termination with a balanced separator in one of the rounds.
We break this goal into three concrete steps.
\underline{(Step 1)} Formulating a (new) continuous optimization problem $F_{\mu,\iota}(\lambda)$ that captures the balanced separators. (For now, the parameters $\mu$ and $\iota$ are not important.) Our formulation also relies on flow-cut duality.
Unlike typical optimization problems solved via FTRL, the variable $\lambda$ in our problem has an exponentially large dimension, and $F_{\mu,\iota}(\lambda)$ is non-convex.
\underline{(Step 2)} Modeling the continuous optimization problem as an online learning problem.
\underline{(Step 3)} Implementing each round of the online learning using the FTRL framework.
As we mentioned in Step 1, our optimization has an exponential number of dimensions, so we cannot simply follow the standard FTRL template, specifically the step of designing the loss function $\ell_t$. Here, we simplify the problem greatly to derive a proxy (and simpler) optimization problem $G_{\mu,\iota}(\lambda)$.
A key technical innovation that we want to highlight here is that we can relate $\nabla G_{\mu,\iota}(\lambda)$,
the gradient of $G_{\mu,\iota}(\lambda)$,
to the (vertex-weighted) graph radius, from which we can implement each FTRL round by doing a simple low-diameter decomposition on the input graphs. We will give a more detailed exposition of these steps in \Cref{subsec:proof-overview} below.

We mentioned above that the linear-time algorithm for minor-free graphs by Bonnet et al.~\cite{bonnet2026separator} is obtained through a novel technique that assigns weights to the graph's vertices. As the algorithm proceeds through $\Theta(h^2)$ iterations, it adjusts the vertices' weights and outputs the separator in one of the iterations. Their vertex-weighting scheme is perhaps the most innovative part of their paper, yet also the most mysterious. Our paper is inspired by this weighting scheme, but we go one step further: by making a connection to the FTRL framework, the weighting scheme arises naturally from Step 3, where we design the loss function $\ell_t$. Our resulting weighting scheme is indeed simpler and more efficient: we can improve the running time from $O(h^{13}n)$~\cite{bonnet2026separator} to $O(h^{7}(m+n))$ without any additional effort, while improving the separator size from $O(h^{13}\sqrt{n})$~\cite{bonnet2026separator} to nearly optimal as in \Cref{thm:minor-separator}.

\subsection{Technical overview}\label{subsec:proof-overview}

We now describe the ideas leading to our algorithm,
which uses a learning game with early termination.

\paragraph{Step 1: Designing an objective.}
Our objective will be inspired by three related concepts:
An existing probabilistic path bound, subset flow problems, and the stochastic connectors
of Bonnet et al.~\cite{bonnet2026separator}.

Our starting point is a key component of Korhonen and Lokshtanov's proof of \cref{thm:KL-separator}.
They gave a bound on a vertex-variant of the $L_\infty$-concurrent flow problem studied by Leighton and Rao~\cite{LR99}, which routes one unit of flow between every pair of vertices, and seeks to minimize the maximum ``congestion'' (sum of flows) through any vertex. The \emph{maximum} congestion is itself called the $L_\infty$ congestion, hence the name ``$L_\infty$-concurrent flow problem''.
This problem had been bounded for many classes in earlier works~\cite{BLR10,matouvsek2014near,lee:LIPIcs.ITCS.2017.1}.
However, almost all these prior bounds were based on various combinatorial methods and were inherently suboptimal as a result.
Korhonen and Lokshtanov instead used a probabilistic construction, essentially proving the following intermediate result.

\begin{restatable}[{adapted from \cite[proofs of Lemmas~4.2--4.4]{KL24}}]{proposition}{inducedpathinteraction}
\label{prop:induced-path-interaction}
Let $H$ be a nonempty graph and $G$ an $H$-induced-minor-free graph.
Sample $u,v$ independently from a probability distribution $\pi$ on $V(G)$,
then sample a simple $u$--$v$ path $P$ from any fixed distribution for that
ordered pair.
If $u=v$, take $P=\{u\}$.
For an independent copy $P'$, we have
\[
 \Pr[\text{$P$ and $P'$ share a vertex or are joined by an edge}]
 \geq\frac{c_1}{|V(H)|+|E(H)|},
\]
where $c_1>0$ is an absolute constant.
\end{restatable}

Later, Le~\cite{le2026separator}
and then Kolbe and Spalding-Jamieson~\cite{kolbe2026eigenvalueboundsboundedgenus}
used variations of this argument for minor-free graphs as well,
essentially proving the following.

\begin{restatable}[{adapted from \cite[proof of Proposition~3.2]{kolbe2026eigenvalueboundsboundedgenus}}]{proposition}{minorpathintersection}
\label{prop:minor-path-intersection}
Let $G$ be a $K_h$-minor-free graph, where $h\geq3$, and let $\pi$ be a
probability distribution on $V(G)$.
Sample $u,v$ independently from $\pi$, then sample a simple $u$--$v$ path $P$
from any fixed distribution for that ordered pair.
If $u=v$, take $P=\{u\}$.
For an independent sample $P'$ resulting from the same process,
\[
 \Pr[V(P)\cap V(P')\neq\varnothing]\geq\frac{c_0}{h^2},
\]
where $c_0>0$ is an absolute constant.
\end{restatable}

Both of these propositions will come up again later in our work,
but for now we will use them only as general ideas.
Korhonen and Lokshtanov and Le used these ideas to bound
the $L_\infty$-concurrent flow problem,
while Kolbe and Spalding-Jamieson
used them to bound the $L_2$-concurrent flow problem
initially studied by Biswal, Lee, and Rao~\cite{BLR10}.
All of these bounds use the propositions in a straightforward manner:
Assume the existence of a concurrent flow
bypassing the claimed bound,
sample $u,v$ from $V$ uniformly,
and then sample a path from $u$ to $v$ using the flow values,
obtaining a probability below the lower bounds in the propositions.

Kolbe and Spalding-Jamieson
also lifted their bound to get a more general bound on the $L_2$-\emph{subset} flow problem
initially studied by Kelner, Lee, Price, and Teng~\cite{kelner2011metric}.
Subset flow problems
(which are dual to ``spreading metric'' problems),
are generally relaxations of balanced separator problems
(see e.g.~\cite{even1999fast}),
so they are a natural tool to try and use
for finding balanced separators.
These problems can be formulated as follows, for a value $1\leq r\leq n$, and various definitions of congestion:
\begin{equation}
\label{eq:subset-flow}
\begin{split}
    \min_{\substack{\lambda\colon\binom{V}{r}\to\real_{\geq0}\\ F_A\ \text{a multicommodity flow in $G$ for every }A\in\binom{V}{r}}}\quad & \operatorname{congestion}\Bigl(\sum_{A\in\binom{V}{r}}\lambda(A)\,F_A\Bigr)\\
    \text{subject to}\quad & \sum_{A\in\binom{V}{r}}\lambda(A)=1,\\
    & F_A[u,v]=\frac{1}{r^2}\qquad\text{for every }A\in\binom{V}{r}\text{ and }u,v\in A,\\
    & F_A[u,v]=0\qquad\text{for every }A\in\binom{V}{r}\text{ and }u,v\in V\text{ with }\{u,v\}\not\subseteq A,
\end{split}
\end{equation}
where $F[u,v]$ denotes the amount of flow that a multicommodity flow $F$ sends between $u$ and $v$.
This is equivalent to the form used by Kelner, Lee, Price, and Teng,
but the form here is much more relevant to our purposes.
In particular, these kinds of flows are quite similar to what we will use,
while the objective we will use will be more directly inspired by the goal of leveraging the propositions.
%At a high-level another way of viewing these problems more directly involves flows within subsets:
%Each $r$-sized subset $A$ is given some demand allocation $\lambda(A)$,
%and

The last ingredient is a slight modification of these subset flows inspired by Bonnet et al.'s ``stochastic connectors''~\cite{bonnet2026separator}.
They define a \defn{stochastic connector} as a collection
of \emph{rooted trees}, each containing a large number of vertices,
so that the collision probability of rooted paths on distinct trees is bounded.
While we will not leverage the specific parametrization of their definition,
it turns out that this general idea is exactly what we need:
Rooted trees can be viewed as a naturally structured variant of the flows $F_A$ in \cref{eq:subset-flow}
that are easy to work with and compute.
The analogous flows here are those between pairs of vertices on the rooted tree, and they are sent only along the tree itself.
Moreover, collision probability is exactly what will allow us to leverage
\cref{prop:induced-path-interaction,prop:minor-path-intersection}.

That said, rooted trees alone are not \emph{quite} the right analogue,
since they are slightly too restrictive.
Instead, we define the idea of a \defn{wide connector}:
A rooted tree \emph{containing} a subset of $3n/4$ specified \defn{terminals}.
The flows we use to sample our paths will then be the root-to-terminal paths,
rather than all rooted paths.

Altogether, this would give us an objective,
defined over wide connectors,
given by the probability of the two different kinds of collisions bounded in
\cref{prop:induced-path-interaction,prop:minor-path-intersection}
(with the two parameters controlling how to weigh the two kinds of collision).
To simplify bookkeeping,
we define our objective
$F_{\mu,\iota}$
as a maximization problem
capturing only the probability of collisions between distinct wide connectors,
while also incorporating a term serving as an analogue to
the total cost sum constraint in
\cref{eq:subset-flow} ($\sum\lambda(A)=1$).
An extra constraint is also required so that our
online learning algorithm is able to make sufficient forward progress
(an analogue of ``width'' in many common learning algorithms).
The precise choice is given
in \cref{subsec:opt-vs-separator}.

\paragraph{Step 2: Formulating an online learning problem.}
We formulate the two players of the learning game as follows.
In each round $t$, the learner plays a vector $\lambda_t$ of weights over all wide connectors,
and the adversary responds with a loss function.
We constrain the game so that, at each round $t$,
$\lambda_t$ has weight updates induced
by at most one new wide connector (in fact, up to normalization only one entry changes),
or the game is terminated early with a small balanced separator. If too many rounds complete
($O(h^2)$ for $K_h$-minor-free graphs, $O(|V(H)|+|E(H)|)$ for $H$-induced-minor-free graphs),
an application of
either
\cref{prop:minor-path-intersection}
or
\cref{prop:induced-path-interaction}
will be used to prove that the graph contains a minor or induced-minor, respectively.

\paragraph{Step 3: Implementing each round of the online learning problem using the FTRL framework.}
It remains to discuss several implementation details.
First, during the course of the algorithm,
we maintain the list of wide connectors $C_t$ played so far,
as well as a \defn{load} vector $p_t$ for each one,
containing the probability of being an element of a sampled path (according to a certain distribution).
We also maintain a prefix sum vector $q_t$ of load vectors.
At each iteration, the adversary uses these values
to define a weight vector $w_t$ over the vertices,
inducing a vertex-weighted shortest-path metric
(equivalently, an edge-weighted shortest-path metric where each edge $uv$ is given weight $\frac12\left(w_t[u]+w_t[v]\right)$).
Using this metric, we will leverage a form of the flow-cut gap for induced-minor-free graphs.
Davies, Rz{\k{a}}{\.z}ewski, Spalding-Jamieson, and Walczak
recently proved the following result.
\begin{theorem}[{\cite[Theorem 3.4]{DRSW27}}]
\label{thm:pure-pairs-subdiv-poly}
Let\/ $G=(V,E)$ be a graph with vertex weights\/ $w\colon V\to\real_{\geq 0}$, $d=d_w^G$ be the vertex-weighted shortest-path metric of\/ $G$, $H$ be a graph with\/ $h_V$ vertices and\/ $h_E\geq 1$ edges, and\/ $\delta,\epsilon,\eta\geq 0$.
If\/ $(2h_V+2)\epsilon+(h_V+h_E)\eta<1$, then, in time polynomial in\/ $|V|$, one can obtain one of the following outputs:
\begin{enumerate}
\item a vertex\/ $x\in V$ with\/ ${|B(x,11\delta)|}>\eta{|V|}$;
\item vertex sets\/ $A,B\subseteq V$ with\/ ${|A|}>\epsilon{|V|}$, ${|B|}>\epsilon{|V|}$, and\/ $d(A,B)>\delta$;
\item an induced minor model of the\/ $1$-subdivision of\/ $H$ in\/ $G$.
\end{enumerate}
\end{theorem}

We give an expected-linear-time (for constant $H$) analogue of this result
for some specific choices of $\delta,\epsilon,\eta$
in \cref{sec:cut-or-cluster}.
The three outcome types induce different behaviors for the adversary:
The first kind of outcome will be sufficient for the adversary to choose a new wide connector
and use it to generate a corresponding loss function;
the second kind of outcome would be needed in an interative manner
(which may instead terminate early with either the first or third kind of outcome),
but would be eventually sufficient for the learner to produce a small separator,
and the third kind of outcome of course gives an induced minor model in $G$.

For the first type of outcome,
a loss function for the learning game is generated using
a proxy objective $G_{\mu,\iota}(p_t)$,
which itself is defined using
a new load vector $p_t$ and the prefix sum of previous load vectors $q_t$.
These vectors, unlike $\lambda_t$, are of linear size,
and are thus much easier to work with computationally.
Our particular choice of $G_{\mu,\iota}$
also gives us a direct relationship between
$\nabla G_{\mu,\iota}(\lambda)$
and the vertex-weighted graph radius.
These are \emph{almost} the weights induced above,
except that we also require a \emph{regularization} term
as part of the FTRL framework,
and some appropriate scaling.
The scaling happens to correspond to the idea of a \emph{learning rate} as defined in various optimization methods,
as well as the diameter of our metric.

Note that the boundary between the learner and adversary
(e.g., which one of them computes the vertex weights or the prefix sum)
becomes fuzzy within the efficient implementation of the algorithm.
In fact, in the actual algorithm we implement,
the weights $\lambda_t$ are not computed explicitly at all.

\section{Preliminaries}

We use mostly standard graph terminology.  Given a graph $H$ and a graph $G$, a \EMPH{minor model} of $H$ in $G$ is a collection $\cK$ of vertex-disjoint connected subgraphs of $G$ such that:
\begin{itemize}
    \item  \textbf{[Vertex correspondence]:} there is a one-to-one correspondence between subgraphs of $\cK$ and vertices of $H$. This means $\cK$ has exactly $|V(H)|$ subgraphs.
    \item   \textbf{[Edge correspondence]:} For every edge $(u,v)\in E(H)$, there is an edge in $G$ between the two corresponding subgraphs $C_u$ and $C_v$ in $\cK$.
\end{itemize}

$G$ is \EMPH{$H$-minor-free} if it does not contain a minor model for $H$. 

We say that $\cK$ is an \EMPH{induced minor model} of $H$ in $G$ if, in addition to satisfying the vertex correspondence and edge correspondence properties above,  it satisfies:
\begin{itemize} 
    \item \textbf{[Non-edge correspondence]:}  If there is no edge between two vertices $u$ and $v$ of $H$, then there is no edge in $G$ between the two corresponding subgraphs $C_u$ and $C_v$ in $\cK$.
\end{itemize}

$G$ is \EMPH{$H$-induced-minor-free} if it does not contain an induced minor model for $H$.  Because of the non-edge correspondence property, an $H$-induced-minor-free graph can be dense. For example, the clique $K_n$ excludes $P_2$ (a path of length 2) as an induced minor.  Every graph $H$ with $|V(H)| = h$ is an induced minor of $K^{(1)}_h$, so if $G$ does not contain $H$ as an induced minor, then it does not contain $K^{(1)}_h$ as an induced minor either. Thus, in some parts of our paper, we work directly with $K^{(1)}_h$-induced-minor-free graphs.

We also use the following single-source shortest path algorithm on graphs with integer weights by Dial~\cite{Dial69}.

\begin{theorem}[Dial's algorithm~\cite{Dial69}]\label{thm:Dial} Let $G = (V,E,w)$ be a graph with integer weights on the edges. Let $D$ be the diameter of $G$, $n = |V|$, and $m = |E|$. Then computing the single-source shortest path from any source vertex takes $O(n+m+D)$ time. 
\end{theorem}

\section{Balanced Separators from Follow the Regularized Leader}

Since our algorithms for balanced separators are inspired by the Follow the Regularized Leader (FTRL) framework in online learning, we first give a brief overview of online learning and FTRL.

\subsection{Online Learning: A Primer}\label{subsec:online-FTRL}

\paragraph{The setup.} The general setup is a game over $T$ rounds, where in each round $t \in [T]$, a player picks a vector $x_t\in \cX$, called a \EMPH{strategy}, where $\cX\subseteq \real^{d}$ is the domain. (Often $\cX$ is a convex set.) The player could choose $x_t$ based on the information from previous rounds. Once the player chooses $x_t$, a loss function $\ell_t: \cX \rightarrow (-\infty,+\infty]$ is revealed; the player pays $\ell_t(x_t)$ loss; and the game continues to the next round.

The goal of the player is to minimize the \EMPH{regret}, defined as follows:
\begin{equation}\label{eq:regret}
    \Reg(x_1,\ldots, x_T) = \sup_{u\in \cX}\left(\sum_{t=1}^T (\ell_t(x_t) - \ell_t(u))\right)
\end{equation}
which is the loss difference w.r.t. the best fixed strategy.

\paragraph{Follow the regularized leader (FTRL)} is a meta-algorithm that a player can use to pick a strategy $x_t\in \cX$ in every step. To apply this meta-algorithm, the player has to first choose a function $\psi: \cX \rightarrow \real$ called a \EMPH{regularizer}. Once the regularizer is chosen, in every step $t$, the player chooses a strategy $x_t\in \cX$ that minimizes $\psi(x) + \sum_{i=1}^{t-1}\ell_i(x)$. That is:
\begin{equation}\label{eq:x_t-choose}
    x_t = \argmin_{x\in \cX}\left(\psi(x) + \sum_{i=1}^{t-1}\ell_i(x)\right)
\end{equation}

The second term in \Cref{eq:x_t-choose} is not hard to see: the only information given up to round $t$ is the loss functions $\{\ell_i(x)\}_{i < t}$, and it makes sense to find a strategy for round $t$ that minimizes the sum $\sum_{i=1}^{t-1}\ell_i(x)$. On the other hand, it is well known in online learning that minimizing this sum alone is not enough, since the adversary who knows the strategy could come up with a new loss next round to throw off the player; see, e.g.,~\cite{Trevisan2019Lecture12}. A simple fix that works well is to add the regularizer $\psi(x)$~\cite{AHR08}.  The algorithm's pseudocode is given in \Cref{alg:FTRL}.

\begin{algorithm}[H]
\caption{$\textsc{FTRL}(\psi,\mathcal X)$: Follow the Regularized Leader Meta-Algorithm}\label{alg:FTRL}
\textbf{Input:} Regularizer $\psi: \cX\rightarrow \real$\\
\textbf{Output:} $\{x_1,x_2,\ldots, x_T\}$
\begin{algorithmic}[1]
   \For{each round $t\leftarrow 1$ to $T$}
        \State pick   $\displaystyle x_t\in\argmin_{x\in\mathcal X}
\left\{\psi(x)+\sum_{i=1}^{t-1}\ell_i(x)\right\}$
        \State loss function $\ell_t(x):\mathbb R^d\to(-\infty,+\infty]$ is revealed
        \State player pays loss $\ell_t(x_t)$
    \EndFor
\State \Return $\{x_1,x_2,\ldots, x_T\}$
\end{algorithmic}
\end{algorithm}

The choice of the regularizer matters in minimizing the regret bounds. We will not discuss the details of these choices, as we will not use these regret bounds in our paper. Curious readers are referred to \cite{orabona2026online} for details. Instead, we briefly review how FTRL solves a convex optimization problem, since our algorithms for balanced separators follow the same strategy. 

\paragraph{Convex optimization via FTRL.} Given a convex function $f(x): \cX \rightarrow \real$ where $\cX \subseteq \real^{d}$ is convex,  find:
\begin{equation}\label{eq:convex-opt}
    \argmin_{x\in \cX}f(x)
\end{equation}
We can approximately solve this problem via the FTRL framework as follows. First, choose any initial feasible point $x_0\in \cX$,  a regularizer $\psi(x) = \frac{\eta}{2}\|x - x_0\|_2^2$ where $\eta > 0$ is a parameter called the \EMPH{learning rate}. Then we play the online game for $T$ rounds (with $T$ chosen later), but for doing so, we need to choose a loss function $\ell_t(x)$ at each round $t\in [T]$. 

We could simply choose $\ell_t(x) = f(x)$, but then $\sum_{i=1}^{t-1}\ell_i(x) = (t-1)f(x)$ and hence finding $x_t$ following \Cref{eq:x_t-choose} is as hard as solving the original convex optimization problem in \Cref{eq:convex-opt}. Instead, we choose $\ell_t(x) = \langle g_t,x\rangle$  where $g_t \in \partial f(x_t)$ is a subgradient of $f$ at $x_t$. Intuitively, $\ell_t(x)$ is a linear approximation of $f(x)$ at time $t$ since by the definition of subgradient:
\begin{equation}
    f(x)\geq f(x_t) +  \langle g_t,x-x_t\rangle
\end{equation}
and hence, instead of minimizing $f(x)$ directly (which is a hard problem), we minimize the linear lower bound. With this choice of $\ell_t(x)$, \Cref{eq:x_t-choose} becomes:
\begin{equation}\label{eq:x_t-convexopt}
    x_t = \argmin_{x\in \cX}\left(\frac{\eta}{2}\|x - x_0\|_2^2 + \langle \sum_{i=1}^{t-1} g_i, x\rangle \right)~,
\end{equation}
which is a simple quadratic function that is much easier to solve.  After round $T$ is completed, we simply return:
\begin{equation}\label{eq:bar-x}
    \Bar{x} = \frac{\sum_{i=1}^{T}x_i}{T}
\end{equation}
as an approximate solution to \Cref{eq:convex-opt}. By choosing $T$ and $\eta$ appropriately, one can show that $f(\Bar{x}) - f(x^*)\leq \epsilon$ for any given $\epsilon \in (0,1)$. Again, we will not reproduce the analysis here, but note two important points that are relevant to us:

\begin{enumerate}
    \item Choosing the regularizer $\psi(x)$ and the proxy loss function $\ell_t(x)$ at each round $t\in [T]$ is important for efficient computation of the strategy $x_t$ at each step, as well as for obtaining a good approximate solution $\Bar{x}$ in the end.
    \item Though the analysis of FTRL works out nicely if $f(x)$ is convex, the algorithm also makes sense when $f$ is non-convex. In the latter case, the kind of guarantee one gets from FTRL depends on how nicely $ f$ ``behaves''. 
    
\end{enumerate}
 In our setup of FTRL for balanced separators, we not only have to choose $\psi(x)$ and $\{\ell_t(x)\}_{t\in [T]}$, but we also have to set up an appropriate function $f$, which turns out to be non-convex and very non-standard. Setting up $f$ is the focus of the next section. 

 \subsection{Balanced Vertex Separators as A Continuous Optimization Problem}\label{subsec:opt-vs-separator}

 A starting point to formulate our optimization problem is the intimate relationship between balanced separators and \EMPH{concurrent flow}.  Let $\mathcal{P}_G$ be the set of all simple paths in an undirected graph $G = (V,E)$ with $n$ vertices and $m$ edges. A \EMPH{concurrent flow} $\lambda: \mathcal{P}_G\rightarrow [0,1]$ assigns each simple path $P\in \mathcal{P}_G$ a flow value $\lambda(P)\in [0,1]$ such that for every ordered pair $(u,v)\in V^2$,  $\sum_{P\in \cP_{uv}}\lambda(P) = 1/n$ where $\cP_{uv}\subseteq \mathcal{P}_G$ is the set of all simple paths from $u$ to $v$. (Note that   $\cP_{uv} = \cP_{vu}$ since the graph is undirected). In words, a concurrent flow $\lambda$ sends $1$ unit of flow from every vertex to all other vertices, equally distributed.   The \EMPH{congestion} at $u$ of the concurrent flow $\lambda$ is $\sum_{P\in \cP_G: u\in P}\lambda(P)$, and the congestion of $\lambda$ is the maximum congestion over all vertices of $V$.   It is well known that:

 \begin{theorem}[Leighton and Rao~\cite{LR99}]\label{thm:Leighton-Rao} For every $\phi \in (0,1)$, either:
 \begin{enumerate}
     \item $G$ admits a balanced separator of size $\phi \cdot n$, or
     \item $G$ has an induced subgraph $H$ on at least $2n/3$ vertices such that $H$ admits a concurrent flow with congestion $O(\log n /\phi)$.
 \end{enumerate}
 \end{theorem}

There are two closely related interpretations of \Cref{thm:Leighton-Rao} that are useful for us:
\begin{enumerate}
    \item If we can send one unit of flow from every vertex to at least $2n/3$ other vertices with little congestion (large $\phi$), then any balanced separator of $G$ must be large.
    \item  Let $\opt_G$ be the size of the optimal balanced separator of $G$ and $\phi  = \frac{\opt_G}{n} \in (0,1]$. Then:
    \begin{equation}\label{eq:flow/congestion}
        \frac{\text{total flow sent}}{\text{congestion}} \approx\frac{n}{\log(n)/\phi} \approx \Tilde{O}(1) \opt_G
    \end{equation}
\end{enumerate}

 \paragraph{The optimization problem.} Instead of sending flow along paths as in the concurrent flow, we take a slightly different perspective: sending one unit of flow from each vertex $u$ to $3n/4$ other vertices along a tree. This gives rise to a \EMPH{wide connector}, a central concept in our formulation. More formally, a \EMPH{wide connector}, denoted by $C \coloneqq (T,r,B)$, is a tuple consisting of: (i) a rooted tree $T$, which is a (not necessarily spanning) subgraph of $G$, (ii) the root $r$ of $T$, and (iii) a subset $B\subseteq V$ of exactly $3n/4$ vertices\footnote{The constant $3/4$ is arbitrary; any constant in $(2/3,1)$ works equally well.} that \underline{belongs to $T$}. We call $B$ the \EMPH{terminals} of the connector $C$.
 
 Given a wide connector $C = (T,r,B)$, we define its load vector $p_C \in [0,1]^{n}$ as follows:
 \begin{equation}\label{eq:load}
     p_C[v] = \frac{|T_v\cap B|}{|B|} \qquad \forall v\in V
 \end{equation}
 where $T_v$ is the subtree of $T$ rooted at $v$. A subtlety is that not every vertex belongs to $T$, and for such a vertex $v$, we set $p_C[v] = 0$.  A useful interpretation of the load vector is: if we send 1 unit of flow from the root $r$ to all vertices in $B$, equally distributed, along the tree $T$, then $p_C[v]$ is the fraction of the flow going through $v$. Clearly, $p_C[r] = 1$ and $p_C[v] \in [0,1]$ for every $v\in V$.

 Let $\cT(G)$ be the set of all rooted trees that are subgraphs of $G$. For every vertex $u\in V$, let $\Cc_u$ be the set of all wide connectors rooted at $ u$. Formally:
 \begin{equation}\label{eq:connector-u}
     \Cc_u = \{(T,u,B): B \in {V\choose 3n/4}, T\in \cT(G) \text{ rooted at $u$ and }  B\subseteq T\}
 \end{equation}
Let $\Cc = \cup_{u\in V} \Cc_u$ be the set of all wide connectors. Let $\mu$ and $\iota$ be two parameters chosen later. The optimization problem is formulated as follows:

\begin{equation}\label{eq:full-opt}
\begin{split}
     \max_{\lambda \in [0,1]^{|\Cc|}}   & F_{\mu,\iota}(\lambda) \coloneqq \sum_{C\in \Cc}\lambda[C] -  \underbrace{\sum_{(C,C')\in {\Cc \choose 2}}\lambda[C]\lambda[C'] \left( \mu\langle p_C,p_{C'}\rangle + \iota \sum_{(u,v)\in E} (p_C[u]p_{C'}[v] + p_C[v]p_{C'}[u] )\right)}_{\text{congestion term}}\\
\text{subject to }\ &
\sum_{C\in\Cc_u}\lambda[C]\leq1
\qquad\text{for every }u\in V.     
\end{split}
\end{equation}

In this optimization problem:
\begin{enumerate}
    \item for each connector $C = (T,r,B)\in \Cc$ rooted at some vertex $r$, the variable $\lambda[C] \in [0,1]$ denotes the amount of flow $r_C$ sends to $B$ along the tree $T$. Thus, the constraint $\sum_{C\in\Cc_u}\lambda[C]\leq1$ means that the total flow sent from vertex $u$ to all other vertices is at most $1$.
    \item $\sum_{C\in \Cc}\lambda[C]$ is the total amount of flow sent from every node of $G$ to all other nodes via the wide connectors.
    \item  The negative term encodes the flow congestion; it takes some effort to really see this, and we will expand on this term below.  The parameters $\mu$ and $\iota$ determine how much we penalize the total flow sent, depending on the graph classes we are interested in. For example, for minor-free graphs, we set $\mu = 1$ and $\iota = 0$, while for induced-minor-free graphs, we set $\mu = \iota = 1$.
\end{enumerate}
Thus, the optimization in \Cref{eq:full-opt} is to maximize the total flow sent through $G$ via wide connectors, while minimizing the congestion.

\begin{remark}
\label{rmk:exponential}
$\lambda \in [0,1]^{|\Cc|}$ has an exponential number of dimensions, and it is unclear how to solve the optimization problem in \Cref{eq:full-opt} in polynomial time; at least, the optimization problem is non-convex.
\end{remark}

Let:
\begin{equation}\label{eq:bilinear}
    H_{\mu,\iota}(p_C,p_{C'}) = \mu\langle p_C,p_{C'}\rangle + \iota \sum_{(u,v)\in E} (p_C[u]p_{C'}[v] + p_C[v]p_{C'}[u] )
\end{equation}

be the bilinear form of the interaction term. Then \Cref{eq:full-opt} can be shortened to:
\begin{equation}\label{eq:full-shorten-opt}
     \begin{split}
     \max_{\lambda \in [0,1]^{|\Cc|}}    F_{\mu,\iota}(\lambda) \coloneqq \sum_{C\in \Cc}\lambda[C] - & \sum_{(C,C')\in {\Cc \choose 2}}\lambda[C]\lambda[C'] H_{\mu,\iota}(p_{C}, p_{C'})  \\
\text{subject to }\ &
\sum_{C\in\Cc_u}\lambda[C]\leq1
\qquad\text{for every }u\in V.     
\end{split}
\end{equation}

This shorter form is useful for much of our analysis, by exploiting the fact that $ H_{\mu,\iota}(p_C,p_{C'})$ is a bilinear form. In the next subsection, we describe how to approximately solve the optimization problem in \Cref{eq:full-opt} using the FTRL framework laid out in \Cref{subsec:online-FTRL}.

\paragraph{On the congestion term.} The congestion term in \Cref{eq:full-opt} is motivated by the probabilistic perspective of wide connectors and the load vectors, inspired by the argument in~\cite{bonnet2026separator}. For a fixed wide connector $C = (T,r,B)$, if we pick a random vertex $b\in B$, then $p_C[v]$ in \Cref{eq:load} is exactly the probability that $v \in T[r,b]$. Consider the following process to sample two paths in $G$, for a feasible $\lambda$ with at least two positive entries: sample two distinct wide connectors $C\neq C'$ with probability proportional to $\lambda[C]\lambda[C']$, and then sample a random root-to-terminal path from each of them, namely $P$ from $C$ and $P'$ from $C'$. Then:

\begin{enumerate}
    \item the probability that $P$ and $P'$ share a vertex is at most:
    \begin{equation*}
       \frac{\sum_{(C,C')\in {\Cc \choose 2}}\sum_{v\in V}(\lambda[C] p_C[v])(\lambda[C'] p_{C'}[v])}{\sum_{(C,C')\in {\Cc \choose 2}}\lambda[C]\lambda[C']}  = \frac{\sum_{(C,C')\in {\Cc \choose 2}}\lambda[C]\lambda[C'] \langle p_C,p_{C'}\rangle }{\sum_{(C,C')\in {\Cc \choose 2}}\lambda[C]\lambda[C']}
    \end{equation*}
     Thus, the term $\sum_{(C,C')\in {\Cc \choose 2}}\lambda[C]\lambda[C'] \mu\langle p_C,p_{C'}\rangle $ in \Cref{eq:full-opt} captures the collision probability of two random paths. In $K_h$-minor-free graphs, the collision probability of random paths is closely related to the size of the clique minor~\cite{bonnet2026separator}. Thus, by penalizing the optimization problem in \Cref{eq:full-opt} with this collision probability, we indirectly encode the structure of $K_h$-minor-free graphs.
     \item  the probability that  $P$ and $P'$ are connected by an edge $(u,v) \in E$, namely that either $u\in P$ and $v\in P'$ or $v\in P$ and $u\in P'$, is at most:
     \begin{equation}
          \begin{split}
              &\frac{\sum_{(C,C')\in {\Cc \choose 2}}\sum_{(u,v)\in E}\left((\lambda[C] p_C[u])(\lambda[C'] p_{C'}[v]) + (\lambda[C] p_C[v])(\lambda[C'] p_{C'}[u])\right) }{\sum_{(C,C')\in {\Cc \choose 2}}\lambda[C]\lambda[C']}\\
              &= \frac{\sum_{(C,C')\in {\Cc \choose 2}}\lambda[C]\lambda[C']  \sum_{(u,v)\in E} (p_C[u]p_{C'}[v] + p_C[v]p_{C'}[u] ) }{\sum_{(C,C')\in {\Cc \choose 2}}\lambda[C]\lambda[C']}
          \end{split}
     \end{equation}
     Thus, the term $\sum_{(u,v)\in E} (p_C[u]p_{C'}[v] + p_C[v]p_{C'}[u] )$ in \Cref{eq:full-opt} captures the probability that two random paths are connected by an edge. This probability is closely related to the size of an induced minor of $G$~\cite{KL24}.
\end{enumerate}

\subsection{FTRL using Low-Diameter Clusters}

Our goal is to solve:
\begin{equation*}
     \max_{\lambda \in [0,1]^{|\Cc|}}    F_{\mu,\iota}(\lambda)  \qquad \text{subject to }\ 
\sum_{C\in \Cc_u}\lambda[C]\leq1
\qquad\text{for every }u\in V
\end{equation*}
where $ F_{\mu,\iota}(\lambda)$ is defined in \Cref{eq:full-opt}.  Following the FTRL framework, in each round $t$ we are going to pick:
\begin{equation*}
    \lambda_t \in \argmax_{\lambda \in [0,1]^{|\Cc|}}    \{\psi(\lambda) + \sum_{i=1}^{t-1}\ell_i(\lambda)\}
\end{equation*}
and a loss $\ell_t(\lambda)$ which is the proxy for $F_{\mu,\iota}(\lambda)$. However, doing so here is particularly non-trivial since the dimension of $\lambda$  is $|\Cc|$, which is exponential.  To get around this issue, we need to simplify $F_{\mu,\iota}(\lambda)$.% , and (2) the term with $\|\cdot\|_{\infty}$ norm in  $F_{\mu,\iota}(\lambda)$ makes it hard to choose $\ell_t(\lambda)$.

\paragraph{Simplifying the problem.~} First, in every step, we choose $\lambda_t$ that has only one non-zero entry; all other entries are 0. This means at step $t$, we will find only one wide connector $C_t$ such that $\lambda_t[C_t] \not= 0$, while $\lambda_t[C] = 0$ for every $C\in \Cc\setminus \{C_t\}$. Second, we are going to set $\lambda_t[C_t] = 1$ for every $t\in [T]$. That means in every round we will send $1$ unit of flow along the tree of $C_t$ from the root. Third, to find $C_t$ at step $t$, we restrict the set $\Cc$ in the definition of $F_{\mu,\iota}(\lambda)$ to $\Cc_t = \{C_1,\ldots, C_{t}\}$ where $C_1,\ldots,C_{t-1}$ are the wide connectors obtained in the previous steps $1,2,\ldots, t-1$. All of the wide connectors are known except $C_{t}$. To simplify the notation, we denote the load vector $p_{C_t}$ of the wide connector $C_t$ by $p_t$ for $t\in [T]$.  With these simplifications, the problem (at step $t$), which is $\argmax_{\lambda \in [0,1]^{|\Cc|}}    F_{\mu,\iota}(\lambda)$, becomes:

\begin{equation}\label{eq:opt-stept-1}
     \begin{split}
    &\quad \argmax_{C_t\in \Cc}  \left(   \sum_{C\in \Cc_t}\lambda[C] -   \sum_{(C,C')\in {\Cc_t \choose 2}}\lambda[C]\lambda[C'] H_{\mu,\iota}(p_{C}, p_{C'})\right) \\
   &=   \argmax_{C_t\in \Cc}   \left(t- \sum_{\{i,j\}\in \binom{[t]}{2}} H_{\mu,\iota}(p_{i}, p_{j})\right) \qquad \text{(since $\lambda[C_i] = 1~\forall 1\leq i\leq t$)}\\
   &= \argmin_{C_t\in \Cc} \left(\sum_{\{i,j\}\in \binom{[t]}{2}} H_{\mu,\iota}(p_{i}, p_{j}) \right)\quad \text{(dropping constant $t$)}\\
   &= \argmin_{C_t\in \Cc} \left(\sum_{\{i,j\}\in \binom{[t]}{2}} \mu\langle p_i,p_{j}\rangle + \iota \sum_{(u,v)\in E} (p_i[u]p_{j}[v] + p_i[v]p_{j}[u])\right) \quad \text{(by \Cref{eq:bilinear})}\\
   &= \argmin_{C_t\in \Cc}  \left(
   \mu \langle \sum_{i=1}^{t-1}p_i,p_t\rangle + \iota  \sum_{1\leq i\leq t-1} \sum_{(u,v)\in E} (p_i[u]p_{t}[v] + p_i[v]p_{t}[u]) \right)\\
   &= \argmin_{C_t\in \Cc} \left(
    \mu \langle q_t,p_t\rangle + \iota  \sum_{(u,v)\in E} (q_t[u]p_{t}[v] + q_t[v]p_{t}[u]) \right) \quad \text{(by defining \EMPH{$q_t \coloneqq \sum_{i=1}^{t-1}p_i$})}\\
   &\coloneqq \argmin_{C_t\in \Cc}\EMPH{$G_{\mu,\iota}(p_t)$} 
\end{split}
\end{equation}

\begin{observation}\label{obs:linear-G} Function $G_{\mu,\iota}(p_t)$ defined in \Cref{eq:opt-stept-1} is a linear function in $p_t$, and the gradient:
\begin{equation}\label{eq:gradient-G}
  \nabla_{p_t}  G_{\mu,\iota}(p_t)[v] = \mu q_t[v] + \iota  \sum_{u\in N_G(v)} q_t[u]
\end{equation}
for every $v\in V$.
\end{observation}

Recall that $p_t$ is the load vector of the (unknown) wide connector $C_t$, and hence the allowable values for $p_t$ form a (finite) discrete set of exponential size. This is the key difficulty in minimizing $G_{\mu,\iota}(p_t)$, even though it is a linear function of $p_t$.  The next step, which arguably is the most innovative step, is to connect optimizing for $G_{\mu,\iota}(p_t)$ with low-diameter clusters.

\paragraph{Connection to low-diameter clusters.~} Note that $G_{\mu,\iota}(p_t)    = \langle \nabla_{p_t}  G_{\mu,\iota}(p_t), p_t\rangle$ since $G_{\mu,\iota}(p_t)$  is linear. The following lemma connects $G_{\mu,\iota}(p_t)$ to the graph diameter.

\begin{lemma}\label{lm:grad-diam} Let $\Tilde{w}\in \real_{\geq 0}^{V}$ be any non-negative vector.  Let $C_t = (X,r,B)$ be a wide connector and $p_t$ be its load vector. Suppose that we assign every vertex $v\in V$ a weight $\Tilde{w}[v]$. Let  $Q$ be a random path of $X$ obtained by sampling a vertex $b\in B$ uniformly at random and returning the $b$-to-root path $X[r,b]$. Then the expected weight of the path is:
\begin{equation}
    \bbE[\Tilde{w}(Q)] = \langle \Tilde{w},p_t\rangle
\end{equation}
As a result, $\langle \Tilde{w},p_t\rangle \leq D$ where $D$ is the vertex-weighted radius of $X$ centered at $r$.
\end{lemma}
\begin{proof}
  By definition of the load vector, $\Pr[v\in X[r,b]] = p_t[v]$. Thus:
   \begin{equation*}
       \bbE[\Tilde{w}(Q)] = \sum_{v\in V(X)}\Pr[v\in Q]\,\Tilde{w}[v] = \langle \Tilde{w},p_t\rangle
   \end{equation*}
as vertices $u$ not in $X$ have $p_t[u] = 0$. 
\end{proof}

By \Cref{lm:grad-diam}, if we assign each vertex $v$ in the tree of the wide connector $C_t$ a weight $\nabla_{p_t}  G_{\mu,\iota}(p_t)[v]$, then $G_{\mu,\iota}(p_t)$, which is $ \langle \nabla_{p_t}  G_{\mu,\iota}(p_t), p_t\rangle$, is the expected weight of a random path in the tree. Thus, to minimize $G_{\mu,\iota}(p_t)$, a natural idea is to: (i) assign every vertex $v$ of $G$ a weight $\nabla_{p_t}  G_{\mu,\iota}(p_t)[v]$, (ii) find a shortest path tree $T$ of $G$ w.r.t. the weight function $\nabla_{p_t}  G_{\mu,\iota}(p_t)$, and (iii) cut off the tree at a small radius in such a way that at least $3n/4$ vertices are still in the tree. (If the tree has small radius, then a random path to the root would have small expected weight, and hence small $G_{\mu,\iota}(p_t)$.) A small radius in (iii) might not exist, and this is where \Cref{thm:cut-or-cluster} below comes into play. It gives either a cluster of more than $3n/4$ vertices with small diameter, or a small balanced separator in $G$. In the latter case, we simply stop the algorithm and return the separator.

For vertex weights $w$ and a path $P$, let $w(P)\coloneqq\sum_{v\in P}w(v)$. Let $K_h^{(1)}$ be the graph obtained from $K_h$ by subdividing every edge once. Every $K_h$-minor-free graph excludes $K_h^{(1)}$ as an induced minor, and so does every $H$-induced-minor-free graph with $h=|V(H)|$.

\begin{restatable}{theorem}{cutorcluster}\label{thm:cut-or-cluster}
Let $h\geq 2$, and let $G=(V,E)$ be a graph with $n$ vertices and $m$ edges that excludes $K_h^{(1)}$ as an induced minor. Let $w:V\to\mathbb{Z}_{\geq 1}$ be vertex weights, $W\coloneqq\sum_{v\in V}w(v)$, and $0<D\leq n$. There is a randomized algorithm $\textsc{CutOrCluster}(G,h,w,D)$ that returns one of the following:
\begin{enumerate}
    \item a \EMPH{cluster}: a set $K\subseteq V$ with $|K|>3n/4$ such that $G[K]$ has vertex-weight diameter at most $D$: every two vertices of $K$ are joined by a path $P$ in $G[K]$ with $w(P)\leq D$, or
    \item a \EMPH{balanced separator}: a set $S\subseteq V$ with $|S|=O(W\log(hW/n)/D)$ such that every connected component of $G-S$ has at most $3n/4$ vertices.
\end{enumerate}
Its expected running time is $O(h^5(\Tsssp(n+1,m+n,2W)+m+n))$, where $\Tsssp(N,M,L)$ is the time to compute a shortest-path tree from a single source in a graph with at most $N$ vertices, at most $M$ edges, and integer edge lengths with maximum distance at most $L$. 
\end{restatable}

We prove \Cref{thm:cut-or-cluster} in \Cref{sec:cut-or-cluster}, building on a structural result of \cite{DRSW27}. We note that while \Cref{thm:cut-or-cluster} is formalized for induced-minor-free graphs, it applies equally well to minor-free graphs, and we actually use it to prove our separator result for minor-free graphs.

In our implementation, instead of assigning the weight exactly $\nabla_{p_t}  G_{\mu,\iota}(p_t)[v]$ to every vertex $v$, we will assign this weight scaled by a diameter parameter $D > 0$; it is $4D$ to be precise. (And we also have to take into account the regularization following the FTRL framework.) Concretely, we assign a weight:
\begin{equation}\label{eq:wt(v)}
    \begin{split}
    w_t[v] &=  (1 + \frac{n\deg(v)}{2m})+ 4D  \nabla_{p_t}  G_{\mu,\iota}(p_t)[v] \\ &= (1 + \frac{n\deg(v)}{2m})
            +4D\left(
                \mu q_t[v]+\iota\sum_{u\in N_G(v)}q_t[u]
            \right)~,
    \end{split}
\end{equation}
for every $v\in V$. The first term is the regularization: it is the gradient of the linear regularizer
\begin{equation}\label{eq:regularizer}
    \psi(\lambda)\coloneqq -\frac{1}{4D}\sum_{C\in\Cc}\lambda[C]\sum_{v\in V}\left(1+\frac{n\deg(v)}{2m}\right)p_C[v]~,
\end{equation}
so for the indicator vector $\lambda_t$ of $C_t$, maximizing $\psi(\lambda_t)$ plus the objective in \Cref{eq:opt-stept-1} is the same as minimizing $\langle w_t,p_t\rangle$. Its role is to make $w_t[v]\geq 1$ with $\sum_{v\in V}(1+\frac{n\deg(v)}{2m})=2n$, which the bounds on $\|w_t\|_1$ use. In convex optimization using gradient descent, $D$ corresponds to the \EMPH{learning rate}; in our discrete setting, it is the diameter. Now we have everything we need for the FTRL algorithm.

\paragraph{The algorithm.~} Let us briefly recap the discussion up to this point. At each step $t$ of the FTRL framework, we find a wide connector $C_t\in \Cc$ that minimizes \Cref{eq:opt-stept-1}.  To do so, we assign every vertex $v\in V$ a weight $w_t[v]$ in \Cref{eq:wt(v)}. We then apply \textsc{CutOrCluster} from \Cref{thm:cut-or-cluster} to $G$ w.r.t. the vertex weight function $\ceil{w_t}$ with diameter parameter $D$. Here $\ceil{w_t}$ is the \EMPH{component-wise ceiling} of $w_t$; that is,  $\ceil{w_t} [v] = \ceil{w_t[v]}$ for every $v\in V$.  If it gives a balanced separator (no connected component contains more than $3n/4$ vertices), we return it. Otherwise, it gives a cluster $K$ of more than $3n/4$ vertices, and we form a wide connector $C_t$ that contains the shortest path tree (with vertex weight $\ceil{w_t}$) of $G[K]$. The pseudocode is given in  \Cref{alg:padded-ftrl} below.

\begin{algorithm}[H]
\caption{$\textsc{Cluster-FTRL}(G,h,T,D,\mu,\iota)$}
\label{alg:padded-ftrl}
\textbf{Input:} $h$ such that $G$ excludes $K_h^{(1)}$ as an induced minor, $T=$ number of FTRL rounds, $D=$ diameter parameter,  $\{\mu,\iota\}=$  parameters in the unified optimization problem in \Cref{eq:full-opt}.\\
\textbf{Output:} either a balanced separator, or a set of wide connectors $\{C_t\}_{t\in [T]T}$, each with $\lambda[C_t] = 1$.
\begin{algorithmic}[1]
    \State $q_1\gets\mathbf{0}$
    \For{$t\gets 1$ to $T$}
        \State $\displaystyle
            w_t[v]\gets
            (1+\frac{n\deg(v)}{2m})
            +4D\left(
                \mu q_t[v]+\iota\sum_{u\in N_G(v)}q_t[u]\right)$ for every $v\in V$ \label{line:wt-def}
        \State run \textsc{CutOrCluster}$(G,h,\ceil{w_t},D)$ \Comment{\lightgray{\Cref{thm:cut-or-cluster}}} \label{line:wt}
        \If{it returns a balanced separator $S_t$}
            \State \Return $S_t$ \Comment{\lightgray{$S_t$ is a balanced separator}}\label{line:St-return}
        \EndIf
        \State $K\leftarrow$ the returned cluster
        \State $r_t\leftarrow$ a vertex in $K$ that is not a root of any existing connector \label{line:root-choice}  \Comment{\lightgray{we  will send $1$ flow from $r_t$}}
        \State $X_t\leftarrow$ vertex-weighted shortest-path tree of $G[K]$ w.r.t.\ $\ceil{w_t}$ rooted at $r_t$.
        \State $B_t\leftarrow$ an arbitrary set of $3n/4$ vertices of $K\setminus \{r_t\}$.
        \State  $C_t \leftarrow (X_t,r_t,B_t)$, and $p_t\leftarrow$  the load vector of $C_t$ \Comment{\lightgray{the wide connector}}
        \State $q_{t+1}\gets q_t+p_t$
    \EndFor
    \State return $\Cc_T \coloneqq \{C_1,C_2,\ldots, C_T\}$
\end{algorithmic}
\end{algorithm}

We use $\ceil{w_t}$ as the weights rather than $w_t$ for efficient shortest-path computation in linear time (for a fixed $h$). If we are willing to tolerate a factor of $\log n$ in the running time, then we can work with  $w_t$ (though the proof is not simpler in any way).

The following lemma gives the lower bound on the optimal value of the optimization problem in \Cref{eq:full-opt} after $T$ rounds of $\textsc{Cluster-FTRL}$.

\begin{lemma}[Cluster-FTRL Progress]\label{lem:padded-ftrl-progress}  Suppose that  \Cref{alg:padded-ftrl} outputs $\Cc_T$ after $T$ rounds.  Let $\Bar{\lambda}$ be a vector such that, for every wide connector $C\in \Cc$:
\begin{equation}
  \Bar{\lambda}[C] =   \begin{cases}
    0 & \text{if } C\not\in \Cc_T \\
    1 & \text{otherwise} 
\end{cases}
\end{equation}
Then $\Bar{\lambda}$ is a feasible solution of the optimization problem in \Cref{eq:full-opt}  with value:
\begin{equation}
    F_{\mu,\iota}(\Bar{\lambda}) \geq  3T/4%\left(3T/4-\kappa R_\psi(T)\right).
\end{equation}
\end{lemma}
\begin{proof}  We note that connectors in $\Cc_T$ are pairwise distinct since they have different roots due to line~\ref{line:root-choice} in \Cref{alg:padded-ftrl}. Recall in \Cref{eq:bilinear}
\begin{equation*}
     H_{\mu,\iota}(p_C,p_{C'}) = \mu\langle p_C,p_{C'}\rangle + \iota \sum_{(u,v)\in E} (p_C[u]p_{C'}[v] + p_C[v]p_{C'}[u] ).
\end{equation*}
Observe that $ H_{\mu,\iota}(p_C,p_{C'})$ is bilinear w.r.t.\ the variables $p_C$ and $p_{C'}$. We write:
\begin{equation}\label{eq:F-barLambda}
    \begin{split}
        F_{\mu,\iota}(\Bar{\lambda}) &=  \sum_{C\in \Cc}\Bar{\lambda}[C] -  \sum_{(C,C')\in {\Cc \choose 2}}\Bar{\lambda}[C]\Bar{\lambda}[C'] H_{\mu,\iota}(p_{C}, p_{C'})\\
        &= T  -  \sum_{(C,C')\in {\Cc_T \choose 2}} H_{\mu,\iota}(p_{C}, p_{C'})\\
        &= T- \sum_{t=1}^{T} \sum_{i=1}^{t-1} H_{\mu,\iota}(p_i,p_t)\\
        &= T - \sum_{t=1}^{T}  H_{\mu,\iota}(\sum_{i=1}^{t-1} p_i,p_t)  \qquad \text{(since $H_{\mu,\iota}$ is bilinear)}\\
         &=   T - \sum_{t=1}^{T}  H_{\mu,\iota}(q_t,p_t)  \qquad\text{(by definition of $q_t$ in \Cref{alg:padded-ftrl})} 
    \end{split}
\end{equation}
Next, we show \Cref{eq:claim-Hqtpt} below, which together with \Cref{eq:F-barLambda}, implies $ F_{\mu,\iota}(\Bar{\lambda})\geq T - T/4 = 3T/4$ as desired.

\begin{equation}\label{eq:claim-Hqtpt}
    H_{\mu,\iota}(q_t,p_t) \leq 1/4
\end{equation}
To see this, observe that by applying \Cref{lm:grad-diam} to $p_t$ and $\ceil{w_t}$, we get $\langle \ceil{w_t}, p_t\rangle \leq D$, giving:
\begin{equation*}
    \langle w_t, p_t\rangle \leq  \langle \ceil{w_t}, p_t\rangle \leq D
\end{equation*}
By definition of $w_t$ in line~\ref{line:wt-def} of \Cref{alg:padded-ftrl}, we have $w_t[v]\geq  4D (\mu q_t[v]+\iota\sum_{u\in N_G(v)}q_t[u])$, and hence:

\begin{equation*}
    \begin{split}
        D\geq \langle w_t, p_t\rangle \geq 4D
                \left(\mu \langle q_t,p_t \rangle +\iota\sum_{uv\in E}(q_t[u]p_t[v] + q_t[v]p_t[u])\right)  = 4D H_{\mu,\iota}(q_t,p_t)
    \end{split}
\end{equation*}
giving  \Cref{eq:claim-Hqtpt}.
\end{proof}

\section{Balanced Separators for Minor-Free Graphs}

In this section, $G$ is a $K_h$-minor-free graph for an integer $h\geq 3$.  We apply the FTRL algorithm (\Cref{alg:padded-ftrl}) with the following parameters.  The number of FTRL rounds is $T = c\cdot h^2$  for a sufficiently large constant $c$ chosen later. We set the diameter parameter $D  = \sqrt{n/T}$ in \Cref{alg:padded-ftrl}.

Our goal here is to show by contradiction that the FTRL algorithm in \Cref{alg:padded-ftrl} \underline{will not} run all $T$ rounds. That is, it will return a balanced separator in line~\ref{line:St-return}. By \Cref{lem:padded-ftrl-progress}, after $T$ rounds, we have found a feasible solution with value at least $3T/4 = 3c h^2/4$.  The following lemma bounds the value of the optimization problem in \Cref{eq:full-opt} by $O(h^2)$. Thus, as long as we choose $c$ sufficiently large, the algorithm must terminate before round $T$.

\begin{lemma}\label{lm:opt-bound-minor} Let $\lambda$ be any feasible solution of the optimization problem in \Cref{eq:full-opt} with $\mu=1$ and $\iota = 0$. If $G$ is $K_h$-minor-free, then:
\begin{equation}
    F_{1,0}(\lambda)  = O(h^2)
\end{equation}
\end{lemma}

We delay the proof of \Cref{lm:opt-bound-minor} to \Cref{subsec:minor-opt-bound}.  Instead, we will continue with the analysis of the FTRL algorithm for minor-free graphs, specifically bounding the size of the output separator. To this end, we bound the $\ell_1$-norm of $w_t$.

\begin{lemma}\label{lm:wt_qt-struct} In every round $\|w_t\|_1 \leq 6n$.
\end{lemma}
\begin{proof}
  Applying \Cref{lm:grad-diam} to $p_t$ and $\ceil{w_t}$, we get $\langle \ceil{w_t}, p_t\rangle \leq D$, giving:
\begin{equation*}
    \langle w_t, p_t\rangle \leq  \langle \ceil{w_t}, p_t\rangle \leq D
\end{equation*}
Since $w_t[v]\geq 1$ for every $v\in V$ by the definition, we have:
\begin{equation}
    \|p_t\|_1 \leq  \langle w_t, p_t\rangle  \leq D
\end{equation}
and hence by the definition of $q_t$:
\begin{equation*}
    \|q_t\|_1 = \| \sum_{i=1}^{t-1} p_i\|_1 \leq (t-1)D \leq TD
\end{equation*}
Now we can bound $\|w_t\|_1 $. By the definition and the fact that $\mu = 1, \iota = 0$, we get:
\begin{equation*}
    \begin{split}
        \|w_t\|_1  &= n + \frac{n\sum_v\deg(v)}{2m} + 4D\|q_t\|_1\\
        &\leq n + n + 4D(TD) = 2n + 4(\sqrt{n/T})^2T \leq 6n
    \end{split}
\end{equation*}
as claimed.
\end{proof}

\minorseparator*
\begin{proof}
By \Cref{lm:opt-bound-minor} and \Cref{lem:padded-ftrl-progress}, by choosing $c$ (in the definition of $T$) sufficiently large, the FTRL algorithm (\Cref{alg:padded-ftrl}) will terminate before $T$ rounds. Therefore, \textsc{CutOrCluster}$(G,h,\ceil{w_t},D)$ returns a separator $S$. Let $W = \sum_{v\in V}\ceil{w_t}[v] \leq n + \sum_{v\in V}w_t[v] \leq 7n$ by \Cref{lm:wt_qt-struct}. By \Cref{thm:cut-or-cluster}, we have
\begin{equation}
    \begin{split}
    |S| &= O\left(\frac{W\log(hW/n)}{D}\right)\\
       &=  O\left(7n \log(7h)\sqrt{T/n}\right)\\
       &= O(h\log h \sqrt{n}) \qquad \text{(since $T =O(h^2)$)}
    \end{split}
\end{equation}
as claimed.
For the running time, we note that the vertex weights are integers of total weight at most $W \leq 7n$. Thus, a single-source shortest path tree can be computed in time $O(n + m + W) = O(n+m)$ by Dial's algorithm in \Cref{thm:Dial}. Then by \Cref{thm:cut-or-cluster}, the total expected running time is $O\left(T\cdot h^5(m+n)\right) = O(h^7(m+n))$. 
\end{proof}

\subsection{Proof of \Cref{lm:opt-bound-minor}}\label{subsec:minor-opt-bound}

We use a known collision probability bound
that follows from a variant of an argument by Korhonen and Lokshtanov
for induced-minor-free graphs~\cite{KL24},
later modified by
Le~\cite{le2026separator} for minor-free graphs,
and further generalized by Kolbe and Spalding-Jamieson~\cite{kolbe2026eigenvalueboundsboundedgenus}
for a slightly more general task.
We specifically use an intermediate form of the construction
in the proof by Kolbe and Spalding-Jamieson.
%We use the following bound on the intersection probability of two independent paths.

\minorpathintersection*

\begin{proof}
The random-path argument in
\cite[proof of Proposition~3.2]{kolbe2026eigenvalueboundsboundedgenus}
(see also \cite[Section~4]{le2026separator}) produces a $K_h$-minor if
\[
 \Pr[V(P)\cap V(P')\neq\varnothing]
 \leq\left(\frac{1}{50h}+\frac{1}{2500h^2}\right)^2.
\]
Since $G$ is $K_h$-minor-free, it follows that
\[
 \Pr[V(P)\cap V(P')\neq\varnothing]
 >\left(\frac{1}{50h}+\frac{1}{2500h^2}\right)^2
 \geq\frac{1}{2500h^2}.
\]
Thus the claim holds with $c_0=1/2500$.
\end{proof}

This is all we will need to prove \cref{lm:opt-bound-minor}.

\begin{proofof}{\Cref{lm:opt-bound-minor}}
The case $\lambda=0$ is immediate, so assume $\lambda\neq0$.
Write $C=(T_C,r_C,B_C)$ for each connector.
Sample two connectors $C_1,C_2$ independently with probabilities $\lambda[C]/\|\lambda\|_1$,
and then sample $z$ uniformly from $B_{C_1}\cap B_{C_2}$.
Let $Q_i$ be the $r_{C_i}$--$z$ path in $T_{C_i}$, for $i=1,2$.
Concatenate $Q_1$ and $Q_2$, erasing cycles, to get a path $P$ with independent,
identically distributed endpoints, where $P=\{r_{C_1}\}$ if $r_{C_1}=r_{C_2}$.

Fix $C_1,C_2$ and $i\in\{1,2\}$.
By the definition of $p_{C_i}[v]$, exactly $|B_{C_i}|\cdot p_{C_i}[v]$ terminals
in $B_{C_i}$ have a root-to-terminal path containing $v$.
Restricting to terminals in $B_{C_1}\cap B_{C_2}$ cannot increase this count.
Since $Q_i=T_{C_i}[r_{C_i},z]$ and $z$ is uniform in that intersection,
counting the possible values of $z$ gives
\[
\begin{aligned}
 \Pr[v\in V(Q_i)\mid C_1,C_2]
 &=\Pr[v\in V(T_{C_i}[r_{C_i},z])\mid C_1,C_2]\\
 &=\frac{\left|\left\{z^*\in B_{C_1}\cap B_{C_2}:
            v\in V(T_{C_i}[r_{C_i},z^*])\right\}\right|}
         {|B_{C_1}\cap B_{C_2}|}\\
 &\leq\frac{|B_{C_i}|\cdot p_{C_i}[v]}{|B_{C_1}\cap B_{C_2}|}
 \leq\frac{3n/4}{n/2}\,p_{C_i}[v]
 =\frac32p_{C_i}[v].
\end{aligned}
\]
%The last inequality uses $|B_{C_i}|=3n/4$ and $|B_{C_1}\cap B_{C_2}|\geq n/2$.

Repeat the construction independently, with connectors $C'_1,C'_2$
and paths $Q'_1,Q'_2,P'$.
Fix $i,j\in\{1,2\}$ and condition on all four connectors.
The two terminal choices are independent, since the constructions use independent
random choices.
For $C_i\neq C'_j$, a union bound
%over shared vertices
gives
\[
\begin{aligned}
 &\Pr[V(Q_i)\cap V(Q'_j)\neq\varnothing\mid C_1,C_2,C'_1,C'_2]\\
 &\quad\leq\sum_v
    \Pr[v\in V(Q_i),\ v\in V(Q'_j)\mid C_1,C_2,C'_1,C'_2]\\
 &\quad=\sum_v
    \Pr[v\in V(Q_i)\mid C_1,C_2]\,
    \Pr[v\in V(Q'_j)\mid C'_1,C'_2]\\
 &\quad\leq\frac94\sum_v p_{C_i}[v]p_{C'_j}[v].
\end{aligned}
\]
%The equality uses independence of the two constructions, and the last inequality
%uses the load bound above for each path.
For $C_i=C'_j$, both paths contain the same root, so this same intersection probability is $1$.
%These bounds depend only on $C_i,C'_j$.
The connectors $C_1,C_2,C_1',C_2'$
%Those connectors
are independent samples from the distribution
$\lambda[C]/\|\lambda\|_1$.
Hence, by using expectation over all four connector choices, we get
\[
\begin{aligned}
 \Pr[V(Q_i)\cap V(Q'_j)\neq\varnothing]
 &=\mathbb E\left[
     \Pr[V(Q_i)\cap V(Q'_j)\neq\varnothing\mid C_1,C_2,C'_1,C'_2]
    \right]\\
 &\leq\sum_C\frac{\lambda[C]^2}{\|\lambda\|_1^2}
 +\sum_{C\neq D}\frac{\lambda[C]\lambda[D]}{\|\lambda\|_1^2}
                   \frac94\langle p_C,p_D\rangle.
\end{aligned}
\]
%Here the first sum accounts for equal connectors and the second for distinct connectors.

%Erasing cycles only removes vertices, so
%$V(P)\subseteq V(Q_1)\cup V(Q_2)$, and likewise for $P'$.
%An intersection between $P$ and $P'$ must therefore occur in at least one of the
%four pairs $(Q_i,Q'_j)$.
Another union bound gives
\[
\begin{aligned}
 \Pr[V(P)\cap V(P')\neq\varnothing]
 &\leq\sum_{i=1}^2\sum_{j=1}^2
       \Pr[V(Q_i)\cap V(Q'_j)\neq\varnothing]\\
 &\leq\frac{4\sum_C\lambda[C]^2+
       9\sum_{C\neq D}\lambda[C]\lambda[D]\langle p_C,p_D\rangle}
      {\|\lambda\|_1^2}.
\end{aligned}
\]

To apply \Cref{prop:minor-path-intersection}, observe that the endpoints of $P$
are $r_{C_1},r_{C_2}$.
In particular, they are independently chosen using the distribution
\[
 \pi(v)=\sum_{C:\,r_C=v}\frac{\lambda[C]}{\|\lambda\|_1}.
\]
%The construction gives a fixed conditional path distribution for each endpoint pair,
%and uses a one-vertex path when the endpoints agree.
Thus $P,P'$ satisfy the hypotheses of \Cref{prop:minor-path-intersection}, giving
\[
 \frac{c_0}{h^2}\leq\Pr[V(P)\cap V(P')\neq\varnothing].
\]
%In the ordered sum over $C\neq D$, each unordered connector pair appears twice.
Using the definition of $F_{1,0}$ and $\lambda[C]^2\leq\lambda[C]$,
the lower and upper bounds therefore yield
\[
 \frac{c_0}{h^2}
 \leq\frac{4\sum_C\lambda[C]^2+
              18(\|\lambda\|_1-F_{1,0}(\lambda))}{\|\lambda\|_1^2}
 \leq\frac{22\|\lambda\|_1-18F_{1,0}(\lambda)}{\|\lambda\|_1^2}.
\]
Rearranging yields
%and maximizing the resulting quadratic yields
\[
 F_{1,0}(\lambda)
 \leq\frac{11}{9}\|\lambda\|_1-\frac{c_0\|\lambda\|_1^2}{18h^2}.
 %\leq\frac{121h^2}{18c_0}
 %=O(h^2).
\]
The right-hand side is a concave quadratic in $\|\lambda\|_1$, maximized at
\[
 \|\lambda\|_1=\frac{11h^2}{c_0}.
\]
Substituting this value gives
\[
 F_{1,0}(\lambda)
 \leq\frac{121h^2}{18c_0}
 =O(h^2).
\]
\aftermath
\end{proofof}

\section{Balanced Separators for Induced-Minor-Free Graphs}

In this section, $G$ is $H$-induced-minor-free for some fixed graph $H$.  We set $\mu = 1, \iota = 1$ in the optimization problem \Cref{eq:full-opt}. We apply the FTRL \Cref{alg:padded-ftrl} with $T = c(|V(H)| + |E(H)|)$  for a sufficiently large constant $c$ chosen later, and $D  = n/\sqrt{mT}$. Similar to the minor-free case, we will show that the FTRL algorithm will terminate before $T$ rounds, by upper bounding the value of the optimal solution. 
\begin{lemma}\label{lm:opt-bound-indminor} Let $\lambda$ be any feasible solution of the optimization problem in \Cref{eq:full-opt} with $\mu=\iota = 1$. If $G$ is $H$-induced-minor-free, then:
\begin{equation}
    F_{1,1}(\lambda)  = O(|V(H)| + |E(H)|)
\end{equation}
\end{lemma}

We delay the proof of \Cref{lm:opt-bound-indminor} to \Cref{subsec:indminor-opt-bound}. Now we bound the $\ell_1$-norm of $w_t$.

\begin{lemma}\label{lm:wt_qt-ind-struct} In every round $\|w_t\|_1 \leq 14n$.
\end{lemma}
\begin{proof}
We only need to tweak the proof of \Cref{lm:wt_qt-struct} slightly. Observe that:
\begin{equation*}
    \langle w_t, p_t\rangle \leq D
\end{equation*}
And since $w_t[v]\geq 1 + \frac{n\deg(v)}{2m}$, the above inequality gives:
\begin{equation*}
    \begin{split}
        \|p_t\|_1 & \leq D\\
         \sum_{v\in V}\frac{n\deg(v)p_t[v]}{2m} &\leq D 
    \end{split}
\end{equation*}
Since $q_t = \sum_{i=1}^{t-1}p_i$, it holds that:
\begin{equation}\label{eq:ind-qt-TD}
    \begin{split}
        \|q_t\|_1  \leq (t-1)D &\leq TD\\
        \sum_{v\in V} \frac{n\deg(v)q_t[v]}{2m} &\leq TD 
    \end{split}
\end{equation}
Now we observe by the definition of $w_t$ that:
\begin{equation}
    \begin{split}
        \|w_t\|_1 &= n + \frac{n\sum_v\deg(v)}{2m} + 4D\|q_t\|_1 + 4D\sum_{v\in V} \sum_{u\in N_G(v)}q_t[u] \\
        &\leq n + n +  4D\|q_t\|_1 + 4D \sum_{v\in V} \deg(v)q_t[v]\\
        & \leq 2n + 4D(TD) + 4D TD \frac{2m}{n} \qquad \text{(by \Cref{eq:ind-qt-TD})}\\
        &=  2n + 4T (n/\sqrt{mT})^2 + 4T (n/\sqrt{mT})^2 \frac{2m}{n}\\
        &\leq 2n + 4n + 8n = 14n 
    \end{split}
\end{equation}
as claimed.
\end{proof}

\indminorseparator*
\begin{proof}

By \Cref{lm:opt-bound-indminor} and \Cref{lem:padded-ftrl-progress}, by choosing $c$  sufficiently large, the FTRL algorithm (\Cref{alg:padded-ftrl}) will terminate before $T$ rounds. Therefore, \textsc{CutOrCluster}$(G,|V(H)|,\ceil{w_t},D)$ returns a separator $S$. Let $W = \sum_{v\in V}\ceil{w_t}[v] \leq n + \sum_{v\in V}w_t[v]$. By \Cref{lm:wt_qt-ind-struct}, $W\leq 15n$. It is well known that if $G$ excludes $H$ as an induced minor, then $G$ excludes $K^{(1)}_{|V(H)|}$ as an induced minor. By \Cref{thm:cut-or-cluster}, we have
\begin{equation}
    \begin{split}
    |S| &= O\left(\frac{W\log(hW/n)}{D}\right)\\
       &=  O\left(15n \log(15|V(H)|)\sqrt{mT}/n\right)\\
       &= O(\log |V(H)| \sqrt{m(|V(H)| + |E(H)|)}) \qquad \text{(since $T =O(|V(H)| + |E(H)|)$)}
    \end{split}
\end{equation}
as claimed.
For the running time, we note that the vertex weights are integers of total weight at most $W \leq 15n$. Thus, a single-source shortest path tree can be computed in time $O(n + m + W) = O(n+m)$ by Dial's algorithm (\Cref{thm:Dial}) and hence by \Cref{thm:cut-or-cluster}, the total expected running time is $O\left(T\cdot |V(H)|^5(m+n)\right) = O(|V(H)|^7(m+n))$. 
\end{proof}

\subsection{Proof of \Cref{lm:opt-bound-indminor}}\label{subsec:indminor-opt-bound}

Similarly to the proof of
\cref{lm:opt-bound-minor},
we will make use of a known collision probability bound.
In this case, the bound follows directly from an argument of Korhonen and Lokshtanov,
rather than a later variant of their construction~\cite{KL24}.

\inducedpathinteraction*
\begin{proof}
The random-path argument in \cite[proofs of Lemmas~4.2--4.4]{KL24}
produces an induced minor model of $H$ if
\[
 \Pr[\text{$P$ and $P'$ share a vertex or are joined by an edge}]
 \leq\frac{c_1}{|V(H)|+|E(H)|},
\]
for some absolute constant $c_1>0$.
Once this probability bound is given, their local-lemma argument only needs
independent vertex images and independent conditional path choices,
so it applies to the endpoint distribution $\pi$.
Since $G$ is $H$-induced-minor-free, it follows that
\[
 \Pr[\text{$P$ and $P'$ share a vertex or are joined by an edge}]
 >\frac{c_1}{|V(H)|+|E(H)|}.
\]
\aftermath
\end{proof}

\begin{proofof}{\Cref{lm:opt-bound-indminor}}
The proof will be similar to that of \cref{lm:opt-bound-minor}.
The case $\lambda=0$ is immediate, so assume $\lambda\neq0$.
Construct $P,P'$ and their root paths $Q_1,Q_2,Q'_1,Q'_2$ exactly as in the
proof of \Cref{lm:opt-bound-minor}. The terminal-counting argument in that proof gives, for $i,j\in\{1,2\}$
and every vertex $v$,
\[
 \Pr[v\in V(Q_i)\mid C_1,C_2]\leq\frac32p_{C_i}[v],
 \qquad
 \Pr[v\in V(Q'_j)\mid C'_1,C'_2]\leq\frac32p_{C'_j}[v].
\]

Fix $i,j\in\{1,2\}$ and condition on all four connectors.
The unprimed and primed terminal choices remain independent.
For any vertices $u,v$, it follows that
\[
\begin{aligned}
 &\Pr[u\in V(Q_i),\ v\in V(Q'_j)\mid C_1,C_2,C'_1,C'_2]\\
 &\quad=\Pr[u\in V(Q_i)\mid C_1,C_2]\,
          \Pr[v\in V(Q'_j)\mid C'_1,C'_2]\\
 &\quad\leq\frac94p_{C_i}[u]p_{C'_j}[v].
\end{aligned}
\]
If $C_i\neq C'_j$, a union bound over shared vertices and both orientations
of each joining edge therefore bounds the conditional probability that
$Q_i,Q'_j$ share a vertex or are joined by an edge by
\[
 \frac94\left(\sum_v p_{C_i}[v]p_{C'_j}[v]
 +\sum_{uv\in E(G)}\left(p_{C_i}[u]p_{C'_j}[v]
                       +p_{C_i}[v]p_{C'_j}[u]\right)\right)
 =\frac94H_{1,1}(p_{C_i},p_{C'_j}),
\]
where the equality is the definition of $H_{1,1}$ in \cref{eq:bilinear}.
If $C_i=C'_j$, both paths contain the same root, so the conditional probability
is $1$.
Averaging over all four connector choices gives
\[
\begin{aligned}
 &\Pr[\text{$Q_i$ and $Q'_j$ share a vertex or are joined by an edge}]\\
 &\quad\leq\sum_C\frac{\lambda[C]^2}{\|\lambda\|_1^2}
 +\frac94\sum_{C\neq D}\frac{\lambda[C]\lambda[D]}{\|\lambda\|_1^2}
                     H_{1,1}(p_C,p_D).
\end{aligned}
\]
Erasing cycles only removes vertices, so any shared vertex or joining edge
between $P,P'$ also occurs for at least one of the four pairs $(Q_i,Q'_j)$.
A union bound over these pairs yields
\[
 \Pr[\text{$P$ and $P'$ share a vertex or are joined by an edge}]
 \leq\frac{4\sum_C\lambda[C]^2+
       9\sum_{C\neq D}\lambda[C]\lambda[D]H_{1,1}(p_C,p_D)}
      {\|\lambda\|_1^2}.
\]

The endpoints of $P$ are $r_{C_1},r_{C_2}$, independently sampled from
\[
 \pi(v)=\sum_{C:\,r_C=v}\frac{\lambda[C]}{\|\lambda\|_1}.
\]
We apply \cref{prop:induced-path-interaction} to $P,P'$ and get the
corresponding lower bound $c_1/(|V(H)|+|E(H)|)$.
%In the ordered sum over $C\neq D$, each unordered connector pair occurs twice.
Using \cref{eq:full-opt} and $\lambda[C]^2\leq\lambda[C]$, we obtain
\[
 \frac{c_1}{|V(H)|+|E(H)|}
 \leq\frac{4\sum_C\lambda[C]^2+
              18(\|\lambda\|_1-F_{1,1}(\lambda))}{\|\lambda\|_1^2}
 \leq\frac{22\|\lambda\|_1-18F_{1,1}(\lambda)}{\|\lambda\|_1^2}.
\]
Rearranging gives
\[
 F_{1,1}(\lambda)
 \leq\frac{11}{9}\|\lambda\|_1
 -\frac{c_1\|\lambda\|_1^2}{18(|V(H)|+|E(H)|)}.
\]
The right-hand side is a concave quadratic in $\|\lambda\|_1$, maximized at
\[
 \|\lambda\|_1=\frac{11(|V(H)|+|E(H)|)}{c_1}.
\]
Substituting this value gives
\[
 F_{1,1}(\lambda)
 \leq\frac{121}{18c_1}\left(|V(H)|+|E(H)|\right)
 =O\left(|V(H)|+|E(H)|\right).
\]
\aftermath
\end{proofof}

\section{Computing a Cluster or a Balanced Separator}\label{sec:cut-or-cluster}

% Shared TikZ styles for all figures of the cut-or-cluster section.
% Colors are from the Okabe--Ito colorblind-safe palette.  Every class of vertices
% and regions is also distinguished by shape, fill or outline, so that the figures
% remain readable in grayscale.
\definecolor{figblue}{RGB}{0,114,178}
\definecolor{figorange}{RGB}{230,159,0}
\definecolor{figverm}{RGB}{213,94,0}
\tikzset{
  figure/.style={x=1cm,y=1cm,line cap=round,line join=round,
    every node/.style={lab,inner sep=1pt}},
  % host graph
  vtx/.style={circle,fill=black!45,inner sep=0pt,minimum size=2.6pt},
  ed/.style={black!35,line width=0.4pt},
  % branch sets (and the prefix Y_t)
  bvtx/.style={circle,fill=black,inner sep=0pt,minimum size=3.6pt},
  bed/.style={black,line width=1.5pt},
  % vertex classes: filled disc = kept, hollow disc = far, small filled disc in the path color = on a connecting path
  cvtx/.style={circle,fill=figblue,draw=figblue!50!black,line width=0.35pt,inner sep=0pt,minimum size=3.4pt},
  fvtx/.style={circle,draw=figorange,fill=white,line width=0.8pt,inner sep=0pt,minimum size=3.6pt},
  svtx/.style={circle,fill=figverm,inner sep=0pt,minimum size=3.2pt},
  ctr/.style={circle,fill=figverm,inner sep=0pt,minimum size=4.4pt},
  smp/.style={diamond,fill=figblue,inner sep=0pt,minimum size=6.5pt},
  % regions
  removed/.style={fill=black!14},                       % B(F,3), the removed region
  near/.style={fill=figblue!14},                        % the ball around the current set
  ring/.style={fill=figblue!40},                        % the new part of that ball
  nearline/.style={draw=figblue,line width=0.5pt},      % contour of the ball
  bound/.style={draw=black,densely dashed,line width=0.6pt}, % contour of a larger ball
  component/.style={draw=black!60,densely dotted,line width=0.6pt}, % a connected component
  branch/.style={fill=black!8},                         % disc behind a branch set
  % connector paths P_ij
  conn/.style={figverm,line width=1.3pt},
  path text/.style={text=figverm},
  % region growing: vertices as discs of diameter w(v), intervals as bars
  disc seed/.style={circle,draw=black,fill=black,text=white,line width=0.6pt,inner sep=0pt,font=\footnotesize},
  disc in/.style={circle,draw=black!60,fill=figblue!32,line width=0.6pt,inner sep=0pt,font=\footnotesize},
  disc boundary/.style={circle,draw=figverm,fill=figblue!32,line width=1.4pt,inner sep=0pt,font=\footnotesize},
  disc out/.style={circle,draw=black!60,fill=white,line width=0.6pt,inner sep=0pt,font=\footnotesize},
  cut/.style={draw=figverm,densely dashed,line width=0.8pt},
  radius/.style={draw=black!70,line width=0.5pt,->},
  interval/.style={draw=black!60,line width=0.5pt},
  interval boundary/.style={draw=figverm,line width=1.2pt},
  volume/.style={fill=figblue!40},
  tick/.style={draw=black,line width=0.7pt},
  % annotations
  lab/.style={font=\small},
  dim/.style={font=\footnotesize},
  panel/.style={font=\small},
  divider/.style={draw=black!15,line width=0.5pt},
  guide/.style={draw=black!40,line width=0.4pt},
  marker/.style={draw=figblue,densely dotted,line width=0.6pt},
  samplepath/.style={draw=figblue,line width=0.9pt},
  axis/.style={draw=black!70,line width=0.6pt,->},
  measure/.style={draw=black!75,line width=0.5pt,<->},
  brace/.style={draw=black!75,line width=0.5pt,decorate,
    decoration={brace,amplitude=3pt}},
  trace/.style={draw=black,line width=1pt},
}

In this section, we prove \Cref{thm:cut-or-cluster}, restated below.

\cutorcluster*

For $X\subseteq V$, let $w(X):=\sum_{v\in X}w(v)$. Following~\cite{DRSW27}, the vertex weight function $w$ induces an edge length function $\ell: E\rightarrow \real_{\geq 0}$ where for each edge $e = (u,v)$, we set $\ell(e) = \frac{w(u) + w(v)}{2}$. The graph distance (for vertex-weighted graphs) in this section is the distance w.r.t. the length function $\ell$, unless explicitly noted otherwise. Thus, a path  $P$ from $u$ to $v$ has length $\ell(P) = w(P)-(w(u)+w(v))/2$. For a vertex-weighted subgraph $H$ of $G$, $X\subseteq V(H)$ and $r\geq 0$, let $B_H(X,r):=\{v\in V(H):d_H(X,v)\leq r\}$.

We will need two results for this theorem. The first result is the following lemma, which is an efficient implementation of \cite[Theorem~3.4]{DRSW27}, whose proof will be delayed to  \Cref{subsec:ball-or-pair}.

\begin{lemma}\label{lem:ball-or-pair}
Let $h\geq 2$, and let $G$ be a  connected $K_h^{(1)}$-induced-minor-free graph with $n$ vertices, $m$ edges and edge lengths in $(0,1]$.  Distances are w.r.t. these edge lengths. Let $b\coloneqq n/(8h^2)$. There is a randomized algorithm, denoted by $\textsc{BallOrPair}(G,h)$, that returns one of the following:
\begin{enumerate}
    \item a vertex $x$ with $|B_G(x,11)|>b$,
    \item a pair $(A,B)$ of vertex-disjoint subsets of vertices, each of size at least $b$  and $d_G(A,B) > 1$. 
\end{enumerate}
It makes $O(h^3)$ shortest-path computations (from sets of sources) and an additional $O(h^3(m+n))$ time, in expectation.
\end{lemma}

Intuitively, \Cref{lem:ball-or-pair} says that either we get a low-diameter cluster containing many vertices, or we have two large sets $A$ and $B$ that are sufficiently far apart.   In the former case, we get a cluster as stated in \Cref{thm:cut-or-cluster}. In the latter case, we do not immediately get the separator. W.l.o.g., we assume that $|B_G(A,1/2)| \leq |B_G(B,1/2)|$ and we call $A$ the \EMPH{smaller side} of the pair $(A,B)$. Then we need one more step: grow the smaller side to find a larger subset $R\supseteq A$ whose boundary, denoted by $\bdry R$, is small. Here \EMPH{$\bdry R$} contains all the vertices in $R$ that have at least one neighbor outside $R$.  The following lemma formalizes this step and follows from a simple region-growing argument of Leighton and Rao~\cite{LR99}.

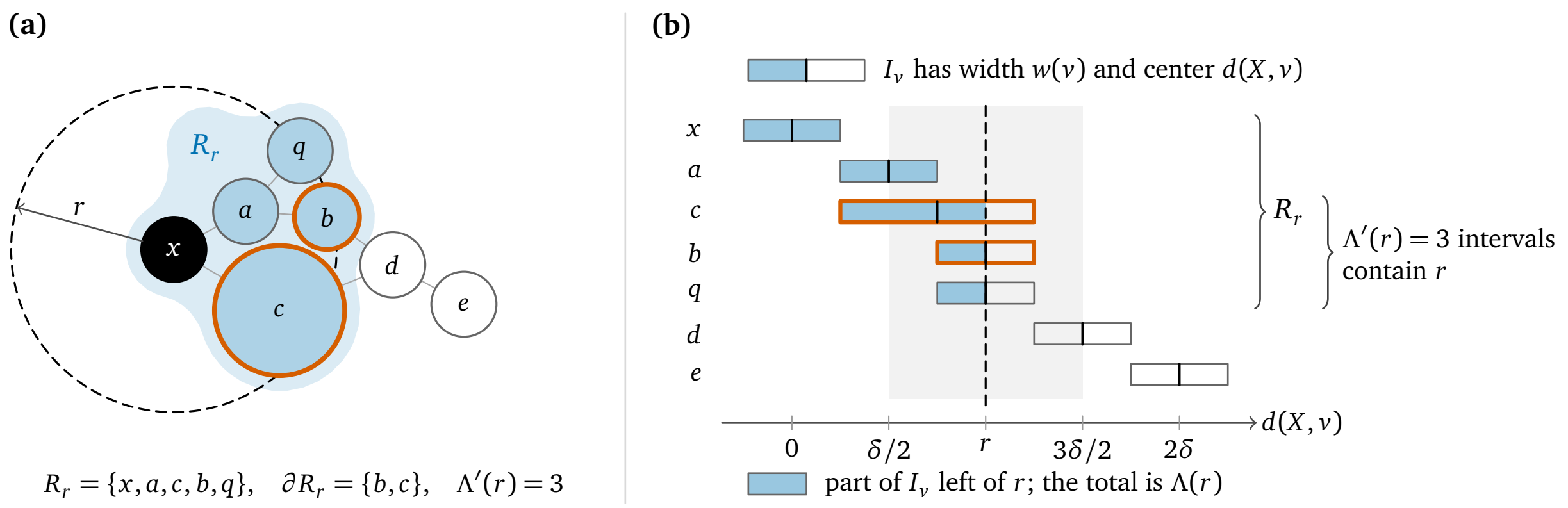
\begin{figure}[h!]
\centering
% Region growing (Lemma lem:region-growing) on one weighted graph: delta = 4, r = 4, w(c) = 4,
% all other weights 2, edge lengths (w(u)+w(v))/2, so d(X,.) = x 0, a 2, c 3, b 4, q 4, d 6, e 8.
% The boundary of R_r is {b,c}; q's interval contains r but q has no neighbour outside R_r.
\begin{tikzpicture}[figure]
\path[use as bounding box] (0,0) rectangle (16.4,5.25);
\draw[divider] (6.46,0.10)--(6.46,5.16);
\node[panel,anchor=west] at (0.04,5.03) {\textbf{(a)}};
\node[panel,anchor=west] at (6.68,5.03) {\textbf{(b)}};
% (a) Each vertex v is a disc of diameter w(v) (slightly shrunk) centred at distance d(X,v) from x,
% so adjacent discs nearly touch, R_r is the set of discs reaching inside radius r, and the
% boundary vertices are among the discs that straddle the circle of radius r.
\path[near] (1.449,3.070)--(1.592,3.192)--(1.658,3.277)--(1.723,3.425)--(1.743,3.585)--(1.723,3.801)--(1.754,3.936)--(1.792,4.007)--(1.835,4.061)--(1.945,4.145)--(2.059,4.186)--(2.168,4.193)--(2.262,4.176)--(2.378,4.128)--(2.483,4.099)--(2.645,4.096)--(2.750,4.120)--(2.961,4.213)--(3.105,4.234)--(3.186,4.227)--(3.262,4.209)--(3.336,4.176)--(3.407,4.131)--(3.488,4.052)--(3.556,3.942)--(3.590,3.836)--(3.604,3.694)--(3.629,3.586)--(3.675,3.484)--(3.783,3.343)--(3.823,3.276)--(3.858,3.186)--(3.874,3.090)--(3.871,2.993)--(3.850,2.898)--(3.816,2.821)--(3.730,2.678)--(3.683,2.523)--(3.677,2.416)--(3.717,2.186)--(3.723,2.105)--(3.707,1.927)--(3.654,1.758)--(3.618,1.685)--(3.543,1.574)--(3.420,1.448)--(3.284,1.357)--(3.134,1.294)--(2.974,1.262)--(2.811,1.262)--(2.651,1.294)--(2.500,1.356)--(2.375,1.438)--(2.293,1.512)--(2.199,1.629)--(2.129,1.758)--(2.058,1.946)--(1.967,2.076)--(1.888,2.146)--(1.797,2.200)--(1.611,2.262)--(1.524,2.308)--(1.426,2.394)--(1.362,2.487)--(1.326,2.575)--(1.308,2.657)--(1.305,2.754)--(1.317,2.831)--(1.342,2.911)--(1.388,2.996)--cycle;
\draw[bound] (1.800,2.720) circle (1.680);
\draw[ed] (1.800,2.720)--(2.542,3.114);
\draw[ed] (2.542,3.114)--(3.380,3.056);
\draw[ed] (2.542,3.114)--(3.104,3.739);
\draw[ed] (1.800,2.720)--(2.891,2.090);
\draw[ed] (2.891,2.090)--(4.059,2.562);
\draw[ed] (3.380,3.056)--(4.059,2.562);
\draw[ed] (4.059,2.562)--(4.794,2.155);
\node[disc seed,minimum size=0.672cm] at (1.800,2.720) {$x$};
\node[disc in,minimum size=0.672cm] at (2.542,3.114) {$a$};
\node[disc boundary,minimum size=1.344cm] at (2.891,2.090) {$c$};
\node[disc boundary,minimum size=0.672cm] at (3.380,3.056) {$b$};
\node[disc in,minimum size=0.672cm] at (3.104,3.739) {$q$};
\node[disc out,minimum size=0.672cm] at (4.059,2.562) {$d$};
\node[disc out,minimum size=0.672cm] at (4.794,2.155) {$e$};
\draw[radius] (1.475,2.807)--(0.177,3.155);
\node[dim,anchor=south] at (0.826,3.041) {$r$};
\node[lab,text=figblue] at (2.142,3.774) {$R_r$};
\node[dim] at (3.150,0.300) {$R_r=\{x,a,c,b,q\}$,\quad $\bdry R_r=\{b,c\}$,\quad $\Lambda'(r)=3$};
% (b) The interval I_v of every vertex; the filled part lies left of r and counts toward Lambda(r).
\begin{scope}[xshift=6.68cm]
\path[fill=black!5] (2.500,1.180) rectangle (4.500,4.200);
\path[volume] (1.000,3.840) rectangle (2.000,4.060);
\draw[interval] (1.000,3.840) rectangle (2.000,4.060);
\draw[tick] (1.500,3.840)--(1.500,4.060);
\node[dim,anchor=east] at (0.620,3.950) {$x$};
\path[volume] (2.000,3.420) rectangle (3.000,3.640);
\draw[interval] (2.000,3.420) rectangle (3.000,3.640);
\draw[tick] (2.500,3.420)--(2.500,3.640);
\node[dim,anchor=east] at (0.620,3.530) {$a$};
\path[volume] (2.000,3.000) rectangle (3.500,3.220);
\draw[interval boundary] (2.000,3.000) rectangle (4.000,3.220);
\draw[tick] (3.000,3.000)--(3.000,3.220);
\node[dim,anchor=east] at (0.620,3.110) {$c$};
\path[volume] (3.000,2.580) rectangle (3.500,2.800);
\draw[interval boundary] (3.000,2.580) rectangle (4.000,2.800);
\draw[tick] (3.500,2.580)--(3.500,2.800);
\node[dim,anchor=east] at (0.620,2.690) {$b$};
\path[volume] (3.000,2.160) rectangle (3.500,2.380);
\draw[interval] (3.000,2.160) rectangle (4.000,2.380);
\draw[tick] (3.500,2.160)--(3.500,2.380);
\node[dim,anchor=east] at (0.620,2.270) {$q$};
\draw[interval] (4.000,1.740) rectangle (5.000,1.960);
\draw[tick] (4.500,1.740)--(4.500,1.960);
\node[dim,anchor=east] at (0.620,1.850) {$d$};
\draw[interval] (5.000,1.320) rectangle (6.000,1.540);
\draw[tick] (5.500,1.320)--(5.500,1.540);
\node[dim,anchor=east] at (0.620,1.430) {$e$};
\draw[bound] (3.500,1.110)--(3.500,4.270);
\draw[axis] (0.78,0.930)--(6.300,0.930) node[dim,anchor=west] {$d(X,v)$};
\draw[guide] (1.500,0.860)--(1.500,1.000);
\node[dim,anchor=north] at (1.500,0.810) {$0$};
\draw[guide] (2.500,0.860)--(2.500,1.000);
\node[dim,anchor=north] at (2.500,0.810) {$\delta/2$};
\draw[guide] (3.500,0.860)--(3.500,1.000);
\node[dim,anchor=north] at (3.500,0.810) {$r$};
\draw[guide] (4.500,0.860)--(4.500,1.000);
\node[dim,anchor=north] at (4.500,0.810) {$3\delta/2$};
\draw[guide] (5.500,0.860)--(5.500,1.000);
\node[dim,anchor=north] at (5.500,0.810) {$2\delta$};
\path[volume] (1.05,4.460) rectangle (1.65,4.680);
\draw[interval] (1.05,4.460) rectangle (2.25,4.680);
\draw[tick] (1.65,4.460)--(1.65,4.680);
\node[dim,anchor=west] at (2.400,4.570) {$I_v$ has width $w(v)$ and center $d(X,v)$};
\path[volume] (1.05,0.190) rectangle (1.65,0.410);
\draw[interval] (1.05,0.190) rectangle (1.65,0.410);
\node[dim,anchor=west] at (1.800,0.300) {part of $I_v$ left of $r$; the total is $\Lambda(r)$};
\draw[brace] (6.280,4.110)--(6.280,2.110);
\node[lab,anchor=west] at (6.440,3.110) {$R_r$};
\draw[brace] (6.980,3.270)--(6.980,2.110);
\node[dim,align=left,anchor=west] at (7.140,2.690) {$\Lambda'(r)=3$ intervals\\contain $r$};
\end{scope}
\end{tikzpicture}%
\caption{Illustrating the region growing algorithm in the proof of \Cref{lem:region-growing}, with $X=\{x\}$, $\delta=4$ and $r=4$. (a) Each vertex $v$ is a disc of diameter proportional to $w(v)$; the boundary vertices of $R_r$ have a thick orange outline. (b) The intervals $I_v$; the filled parts sum to $\Lambda(r)$.}
\label{fig:region-growing}
\end{figure}

\begin{lemma}\label{lem:region-growing}
Let $\delta\geq 1$ be a parameter and let $G = (V,E)$ be a vertex-weighted graph with $n$ vertices, $m$ edges and integer vertex weights $w(v) \in [1,\delta]$. Let $X\subseteq V$ be nonempty.  There is a set $R$ with $X\subseteq R\subseteq B_G(X,2\delta)$ such that  $|\bdry R|\leq \frac{1}{\delta}\ln\frac{w(V)}{w(X)}\cdot w(R)$.  Furthermore, if $\{d_G(X,v)\}_{v\in V}$ is given, then $R$ can be found in $O(m+n+\delta)$ time, and in this case we denote the algorithm by $\textsc{Grow}(G,X,\delta,\{d_G(X,v)\}_{v\in V})$.
\end{lemma}
\begin{proof}
For each vertex $v$, represent it by an interval $I_v = [d_G(X,v)-w(v)/2, d_G(X,v)+w(v)/2]$, which is centered at $d_G(X,v)$ and has width $w(v)$. Then for any edge $e = (u,v)$, $I_v\cap I_u \not=\varnothing$ since $\ell(e) = \frac{w(u) + w(v)}{2}$.  Let $r \in [\delta/2,3\delta/2]$ be a parameter. Let $\Lambda(r)$ be the total length of the parts of the intervals to the left of $r$; see \Cref{fig:region-growing}. Formally:
\begin{equation}
    \Lambda(r) = \sum_{v\in V}\max\{\min\{r,d_G(X,v) + w(v)/2\} -( d_G(X,v) - w(v)/2),0\}
\end{equation}
Let $R_r$ be the set of vertices whose intervals start before $r$.  We observe that:
\begin{enumerate}
    \item $\Lambda'(r)$ is the number of intervals containing $r$ when $r$ is not the endpoint of an interval. (And there are only finitely many such points, whose total measure is $0$ in $[\delta/2,3\delta/2]$.)
    \item  Every vertex $u\in R_r$ such that $I_u$ ends before $r$ only has neighbors inside $R_r$. Thus, every $u\in \bdry R_r$ must have $r\in I_u$, which implies  $|\bdry R_r|\leq \Lambda'(r)$.
    \item $\Lambda(r) \leq w(R_r)$.  
\end{enumerate}

Finally, we observe that:
\begin{equation*}
    \begin{split}
        \int_{\delta/2}^{3\delta/2}\frac{\Lambda'(r)}{\Lambda(r)}\,dr &= \ln(\Lambda(3\delta/2)) - \ln(\Lambda(\delta/2)) \leq \ln(w(V)) - \ln(w(X))  = \ln\frac{w(V)}{w(X)}.
        \end{split}
\end{equation*}
Thus, there exists $r\in [\delta/2,3\delta/2]$ such that $\Lambda'(r) \leq \Lambda(r)\frac{1}{\delta} \ln\frac{w(V)}{w(X)}$, giving:
\begin{equation}
    |\bdry R_r| \leq \Lambda'(r) \leq \Lambda(r)\frac{1}{\delta} \ln\frac{w(V)}{w(X)}\leq \frac{1}{\delta} \ln\frac{w(V)}{w(X)} w(R_r)
\end{equation}
as claimed.  To find such $r$, we simply sweep over endpoints in $[\delta/2,3\delta/2]$ sorted in increasing order, using bucket sort. The running time is $O(n+m+\delta)$.  
\end{proof}

\begin{algorithm}[H]
\caption{$\textsc{CutOrCluster}(G,h,w,D)$}\label{alg:cut-or-cluster}
\begin{algorithmic}[1]
    \State $\delta:=D/25$
    \State $S\leftarrow\{v\in V:w(v)>\delta/4\}$ \label{line:heavy-to-S}
    \State $H\leftarrow$ largest connected component of $G\setminus S$
    \While{$|V(H)| > 3n/4$}
        \State $(x,A,B)\leftarrow \textsc{BallOrPair}(H,h)$ with edge lengths divided by $\delta$.  \Comment{\lightgray{if $x=\textsc{Null}$, then $A$ is the smaller side}}
        \If{$x\not= \textsc{Null}$}
            \If{$|B_H(x,12\delta)|>3n/4$}
                \State \Return $B_H(x,12\delta)$  \Comment{\lightgray{return a cluster}}
            \EndIf
            \State $X\leftarrow B_H(x,11\delta)$
        \Else \Comment{\lightgray{it returns two sets}}
            \State $X\leftarrow A$ \Comment{\lightgray{$|B_H(A,\delta/2)| \leq |B_H(B,\delta/2)| $ and $d_H(A,B) > \delta$}}
        \EndIf
         \State  compute $\{d_H(X,v)\}_{v\in V}$
        \State $R\leftarrow$ \textsc{Grow}$(H,X, \delta/4, \{d_H(X,v)\}_{v\in V})$ \label{line:grow-X}\Comment{\lightgray{apply the algorithm in \Cref{lem:region-growing}} } 
        \State $S\leftarrow S\cup\bdry_H(R)$ \label{line:add-S-sep}
        \State $H\leftarrow$ largest connected component of $H\setminus \bdry_H(R)$  \Comment{\lightgray{also is the largest connected component of $G\setminus S$}}
    \EndWhile
    \State \Return $S$  \Comment{\lightgray{return a cut}}
\end{algorithmic}
\end{algorithm}

Now we are ready to prove \Cref{thm:cut-or-cluster}.  We write \EMPH{$\bdry_H R$} to denote the boundary of $R$ in a graph $H$.

\begin{proofof}{\Cref{thm:cut-or-cluster}} \Cref{alg:cut-or-cluster} implements $\textsc{CutOrCluster}(G,h,w,D)$.  We now describe the intuition of the algorithm.  Let $\delta = D/25$.  First, we include all vertices of weight at least $\delta/4$ in a set $S$, our intended balanced separator. This ensures that all edges have length at most $\delta/4$. The algorithm is iterative, starting from the largest connected component $H$ of $G\setminus S$. If at any point of the algorithm $H$ contains at most $3n/4$ vertices, then we know that $S$ is balanced.

In each iteration, we apply $\textsc{BallOrPair}(H,h)$ (in \Cref{lem:ball-or-pair}) with edge lengths divided by $\delta$, so that every edge has length in $(0,1]$. By \Cref{lem:ball-or-pair}, we either get:
\begin{enumerate}
    \item a vertex $x$ such that (in the unscaled edge lengths),  $B_H(x,11\delta)$ contains more than $|V(H)|/(8h^2)$ vertices. Then we check if $B_H(x,12\delta)$ contains more than $3n/4$ vertices, then we return $B_H(x,12\delta)$, which is a cluster satisfying item 1 of \Cref{thm:cut-or-cluster} since the diameter of $G[B_H(x,12\delta)]$ is at most $2(12\delta) + 2(\delta/4) < 25\delta = D$. Otherwise, we have found a set $X = B_H(x,11\delta)$; we clarify the role of $X$ below. We note that:
    \begin{equation}
        |B_H(X,\delta/2)| \leq |B_H(x,12\delta)| \leq 3n/4
    \end{equation}
    
    \item  a pair $(A,B)$ where $A$ is the smaller side, i.e., $|B_H(A,\delta/2)| \leq |B_H(B,\delta/2)|$. In this case, we take $X = A$. Since $d_H(A,B)>\delta$, the sets $B_H(A,\delta/2)$ and $B_H(B,\delta/2)$ are disjoint, and since $A$ is the smaller side, $|B_H(X,\delta/2)| = |B_H(A,\delta/2)|\leq |V(H)|/2$.
\end{enumerate}

In both cases, $X$ has the following properties: (i) $X$ contains at least $|V(H)|/(8h^2)$ vertices, and (ii) $|B_H(X,\delta/2)|\leq 3n/4$. This means if we grow $X$ by a radius at most $\delta/2$ to get a set $R \subseteq B_H(X,\delta/2)$ by applying \Cref{lem:region-growing} with parameter $\delta/4$ (line~\ref{line:grow-X} in \Cref{alg:cut-or-cluster}), it is guaranteed that if the largest connected component $C$ of $H\setminus \partial_H(R)$ has size more than $3n/4$, then $C\cap R = \varnothing$. Thus, by adding $\bdry_H(R)$ to $S$, in the next iteration we continue with the (largest) connected component of $H\setminus \partial_H(R)$ that is disjoint from $R$. Since $w(X)\geq |X| \geq |V(H)|/(8h^2) > 3n/(32h^2)$, by \Cref{lem:region-growing}, each step we add to $S$ at most:
\begin{equation}
    \frac{4}{\delta}\ln\frac{W}{3n/(32h^2)} w(R) = \frac{4w(R)}{\delta}\ln\frac{32h^2W}{3n}.
\end{equation}
Over the course of the algorithm, the total number of vertices added to $S$ in line~\ref{line:add-S-sep} is:

\begin{equation}
    \sum_R \frac{4w(R)}{\delta}\ln\frac{32h^2W}{3n} \leq \frac{4W}{\delta}\ln\frac{32h^2W}{3n}
\end{equation}

Since in line~\ref{line:heavy-to-S}, we add at most $4W/\delta$ vertices to $S$, the total size of $S$ is:

\begin{equation*}
    |S| \leq \frac{4W}{\delta} +  \frac{4W}{\delta}\ln\frac{32h^2W}{3n}  = O(W\log(hW/n)/D).
\end{equation*}

Now we bound the running time of each iteration.  By \Cref{lem:ball-or-pair}, each $\textsc{BallOrPair}(H,h)$  requires $O(h^3)$ shortest path computations in expectation. Note that in \Cref{alg:cut-or-cluster}, we scale edge lengths by $\delta$ to match the statement of  \Cref{lem:ball-or-pair}, but we will run the single-source shortest path on unscaled lengths. Also note that since each edge $e = (u,v)$ has length $\ell(e) = \frac{w(u) + w(v)}{2}$, which could be fractional, by scaling the edge length by a factor of 2, all the edge lengths are integers. Since the total vertex weight is at most $W$, the maximum distance is at most $2W$. Hence, the total running time of this procedure is $O(h^3(\Tsssp(n+1,m+n,2W)+m+n))$.

To compute $\{d_H(X,v)\}_{v\in V}$, we can reduce to single-source shortest path by adding a dummy vertex $x$ and connecting $x$ to every vertex in $X$ with the same edge weight, say $1$. The running time is $O(\Tsssp(n+1,m+n,2W))$.   Finally, computing $R$ takes $O(n+m+W)$ time by \Cref{lem:region-growing}. Thus, the maximum time per iteration is  $O(h^3(\Tsssp(n+1,m+n,2W)+m+n))$.

We note that since $|X|\geq |V(H)|/(8h^2) > 3n/(32h^2)$, each iteration eliminates more than $3n/(32h^2)$ vertices from $H$. Thus, the total number of iterations is at most $O(h^2)$, and hence the total running time is $O(h^5(\Tsssp(n+1,m+n,2W)+m+n))$, as claimed.
\end{proofof}

\subsection{Proof of \Cref{lem:ball-or-pair}}\label{subsec:ball-or-pair}

 Since $G$ is fixed throughout this subsection, we simplify notation by setting $d:=d_G$ and $B:=B_G$.  Recall that $b = n/(8h^2)$ and that edge lengths are in $(0,1]$.

\begin{figure}[t]
\centering
% FindPrefix. (a) is drawn with unit edges of length u, and every ball of radius r is the
% Euclidean (r+1/2)u-neighbourhood of its center set. DFS order y_1,...,y_t,...; t=4.
\begin{tikzpicture}[figure]
\path[use as bounding box] (0,0) rectangle (16.4,5.25);
\draw[divider] (8.48,0.10)--(8.48,5.16);
\node[panel,anchor=west] at (0.04,5.03) {\textbf{(a)}};
\node[panel,anchor=west] at (8.700000000000001,5.03) {\textbf{(b)}};
% (a) B(Y_{t-1},4) (light), its new part B(Y_t,4) - B(Y_{t-1},4) (dark) and B(y_t,4) (dashed).
\path[near] (3.820,4.570)--(4.001,4.560)--(4.180,4.529)--(4.355,4.479)--(4.523,4.410)--(4.682,4.322)--(4.830,4.217)--(4.966,4.096)--(5.087,3.960)--(5.192,3.812)--(5.280,3.653)--(5.349,3.485)--(5.399,3.310)--(5.430,3.131)--(5.440,2.950)--(5.430,2.769)--(5.399,2.590)--(5.349,2.415)--(5.280,2.247)--(5.192,2.088)--(5.087,1.940)--(4.966,1.804)--(4.830,1.683)--(4.682,1.578)--(4.523,1.490)--(4.355,1.421)--(4.180,1.371)--(4.001,1.340)--(3.820,1.330)--(3.100,1.330)--(2.919,1.340)--(2.740,1.371)--(2.565,1.421)--(2.397,1.490)--(2.238,1.578)--(2.090,1.683)--(1.954,1.804)--(1.833,1.940)--(1.728,2.088)--(1.640,2.247)--(1.571,2.415)--(1.521,2.590)--(1.490,2.769)--(1.480,2.950)--(1.490,3.131)--(1.521,3.310)--(1.571,3.485)--(1.640,3.653)--(1.728,3.812)--(1.833,3.960)--(1.954,4.096)--(2.090,4.217)--(2.238,4.322)--(2.397,4.410)--(2.565,4.479)--(2.740,4.529)--(2.919,4.560)--(3.100,4.570)--cycle;
\path[ring] (4.361,4.560)--(4.540,4.529)--(4.715,4.479)--(4.883,4.410)--(5.042,4.322)--(5.190,4.217)--(5.326,4.096)--(5.447,3.960)--(5.552,3.812)--(5.640,3.653)--(5.709,3.485)--(5.759,3.310)--(5.790,3.131)--(5.800,2.950)--(5.790,2.769)--(5.759,2.590)--(5.709,2.415)--(5.640,2.247)--(5.552,2.088)--(5.447,1.940)--(5.326,1.804)--(5.190,1.683)--(5.042,1.578)--(4.883,1.490)--(4.715,1.421)--(4.540,1.371)--(4.361,1.340)--(4.180,1.330)--(3.820,1.330)--(4.001,1.340)--(4.180,1.371)--(4.355,1.421)--(4.523,1.490)--(4.682,1.578)--(4.830,1.683)--(4.966,1.804)--(5.087,1.940)--(5.192,2.088)--(5.280,2.247)--(5.349,2.415)--(5.399,2.590)--(5.430,2.769)--(5.440,2.950)--(5.430,3.131)--(5.399,3.310)--(5.349,3.485)--(5.280,3.653)--(5.192,3.812)--(5.087,3.960)--(4.966,4.096)--(4.830,4.217)--(4.682,4.322)--(4.523,4.410)--(4.355,4.479)--(4.180,4.529)--(4.001,4.560)--(3.820,4.570)--(4.180,4.570)--cycle;
\draw[nearline] (4.180,4.570)--(4.361,4.560)--(4.540,4.529)--(4.715,4.479)--(4.883,4.410)--(5.042,4.322)--(5.190,4.217)--(5.326,4.096)--(5.447,3.960)--(5.552,3.812)--(5.640,3.653)--(5.709,3.485)--(5.759,3.310)--(5.790,3.131)--(5.800,2.950)--(5.790,2.769)--(5.759,2.590)--(5.709,2.415)--(5.640,2.247)--(5.552,2.088)--(5.447,1.940)--(5.326,1.804)--(5.190,1.683)--(5.042,1.578)--(4.883,1.490)--(4.715,1.421)--(4.540,1.371)--(4.361,1.340)--(4.180,1.330)--(3.100,1.330)--(2.919,1.340)--(2.740,1.371)--(2.565,1.421)--(2.397,1.490)--(2.238,1.578)--(2.090,1.683)--(1.954,1.804)--(1.833,1.940)--(1.728,2.088)--(1.640,2.247)--(1.571,2.415)--(1.521,2.590)--(1.490,2.769)--(1.480,2.950)--(1.490,3.131)--(1.521,3.310)--(1.571,3.485)--(1.640,3.653)--(1.728,3.812)--(1.833,3.960)--(1.954,4.096)--(2.090,4.217)--(2.238,4.322)--(2.397,4.410)--(2.565,4.479)--(2.740,4.529)--(2.919,4.560)--(3.100,4.570)--cycle;
\draw[bound] (4.180,2.950) circle (1.620);
\draw[ed] (3.100,2.950)--(2.788,2.770)--(2.476,2.590)--(2.165,2.410)--(1.853,2.230);
\draw[ed] (3.460,2.950)--(3.460,3.310)--(3.460,3.670);
\draw[ed] (3.820,2.950)--(3.913,2.602)--(4.006,2.255)--(4.100,1.907)--(4.193,1.559);
\draw[ed] (4.180,2.950)--(4.386,3.245)--(4.593,3.540)--(4.799,3.835);
\draw[ed] (2.476,2.590)--(2.414,2.235)--(2.351,1.881);
\draw[ed] (4.006,2.255)--(4.345,2.131)--(4.683,2.008);
\draw[ed] (5.485,2.341)--(5.637,2.015)--(5.789,1.689);
\draw[samplepath] (4.180,2.950)--(4.506,2.798)--(4.833,2.646)--(5.159,2.494)--(5.485,2.341);
\draw[bed] (3.100,2.950)--(3.460,2.950)--(3.820,2.950)--(4.180,2.950);
\node[vtx] at (2.788,2.770) {};
\node[vtx] at (2.476,2.590) {};
\node[vtx] at (2.165,2.410) {};
\node[vtx] at (1.853,2.230) {};
\node[vtx] at (3.460,3.310) {};
\node[vtx] at (3.460,3.670) {};
\node[vtx] at (3.913,2.602) {};
\node[vtx] at (4.006,2.255) {};
\node[vtx] at (4.100,1.907) {};
\node[vtx] at (4.193,1.559) {};
\node[vtx] at (4.386,3.245) {};
\node[vtx] at (4.593,3.540) {};
\node[vtx] at (4.799,3.835) {};
\node[vtx] at (2.414,2.235) {};
\node[vtx] at (2.351,1.881) {};
\node[vtx] at (4.345,2.131) {};
\node[vtx] at (4.683,2.008) {};
\node[vtx] at (4.506,2.798) {};
\node[vtx] at (4.833,2.646) {};
\node[vtx] at (5.159,2.494) {};
\node[vtx] at (5.637,2.015) {};
\node[vtx] at (5.789,1.689) {};
\node[bvtx] at (3.100,2.950) {};
\node[bvtx] at (3.460,2.950) {};
\node[bvtx] at (3.820,2.950) {};
\node[bvtx] at (4.180,2.950) {};
\node[smp] at (5.485,2.341) {};
\node[dim,anchor=south] at (3.100,3.070) {$y_1$};
\node[dim,anchor=south] at (4.100,3.070) {$y_t$};
\node[dim,anchor=north] at (3.640,2.820) {$Y_t$};
\node[dim,anchor=north east] at (5.425,2.241) {$x$};
\node[lab,anchor=east] at (1.380,3.500) {$B(Y_{t-1},4)$};
\node[lab,anchor=south west] at (5.346,4.176) {$B(y_t,4)$};
\draw[nearline] (5.157,1.786)--(5.607,1.436);
\node[dim,anchor=north west] at (5.637,1.406) {$B(Y_t,4)\setminus B(Y_{t-1},4)$};
\node[dim] at (4.150,0.280) {$d(x,y_t)\le 4$,\quad $B(Y_t,4)\setminus B(Y_{t-1},4)\subseteq B(y_t,4)\subseteq B(x,11)$};
% (b) |B(Y_s,4)| as a function of s; the band between b and 2b holds more than b vertices.
\begin{scope}[xshift=8.700000000000001cm]
\path[near] (0.77,1.60) rectangle (5.64,2.95);
\draw[axis] (0.77,0.82)--(5.86,0.82) node[dim,anchor=west] {$s$};
\draw[axis] (0.77,0.82)--(0.77,4.41) node[dim,anchor=south] {$|B(Y_s,4)|$};
\foreach \yy/\ll in {1.72/b,2.62/2b,3.52/3b}{
 \draw[guide] (0.77,\yy)--(5.64,\yy);
 \node[dim,anchor=east] at (0.62,\yy) {$\ll$};
}
\draw[marker] (2.20,0.82)--(2.20,1.60)--(5.79,1.60);
\draw[marker] (4.20,0.82)--(4.20,2.95)--(5.79,2.95);
\draw[trace] (0.77,0.82)--(1.20,0.82)--(1.20,1.08)--(1.70,1.08)
 --(1.70,1.34)--(2.20,1.34)--(2.20,1.60)--(2.70,1.60)
 --(2.70,1.95)--(3.20,1.95)--(3.20,2.18)--(3.70,2.18)
 --(3.70,2.48)--(4.20,2.48)--(4.20,2.95)--(4.70,2.95)
 --(4.70,3.35)--(5.20,3.35)--(5.20,3.91);
\foreach \xx/\yy in {1.20/1.08,1.70/1.34,2.20/1.60,2.70/1.95,3.20/2.18,3.70/2.48,4.20/2.95,4.70/3.35,5.20/3.91}{\node[bvtx] at (\xx,\yy) {};}
\node[dim,anchor=north] at (2.20,0.67) {$i$};
\node[dim,anchor=north] at (4.20,0.67) {$j$};
\node[dim,anchor=north] at (5.20,0.67) {$k$};
\draw[brace] (5.86,2.95)--(5.86,1.60);
\node[dim,align=center,anchor=west] at (5.98,2.28) {$|B(Y_j,4)\setminus{}$\\$B(Y_i,4)|>b$};
\end{scope}
\end{tikzpicture}%
\caption{\textsc{FindPrefix} (\Cref{lm:find-prefix}). (a) A prefix $Y_t=\{y_1,\ldots,y_t\}$ of the depth-first search order and the ball $B(Y_t,4)$. A sampled vertex $x\in B(Y_j,4)\setminus B(Y_i,4)$ is within distance $4$ of some $y_t$ with $i<t\leq j$, and $B(y_t,4)\subseteq B(x,11)$, so if $|B(x,11)|\leq b$ then $|B(Y_t,4)|\leq|B(Y_{t-1},4)|+b\leq 3b$. (b) The size of $B(Y_s,4)$ as a function of $s$: $i$ is the largest index with at most $b$ vertices and $j$ the smallest index with more than $2b$, so the band between them contains more than $b$ vertices, and a uniform sample lands in it with probability at least $1/(8h^2)$.}
\label{fig:find-prefix}
\end{figure}
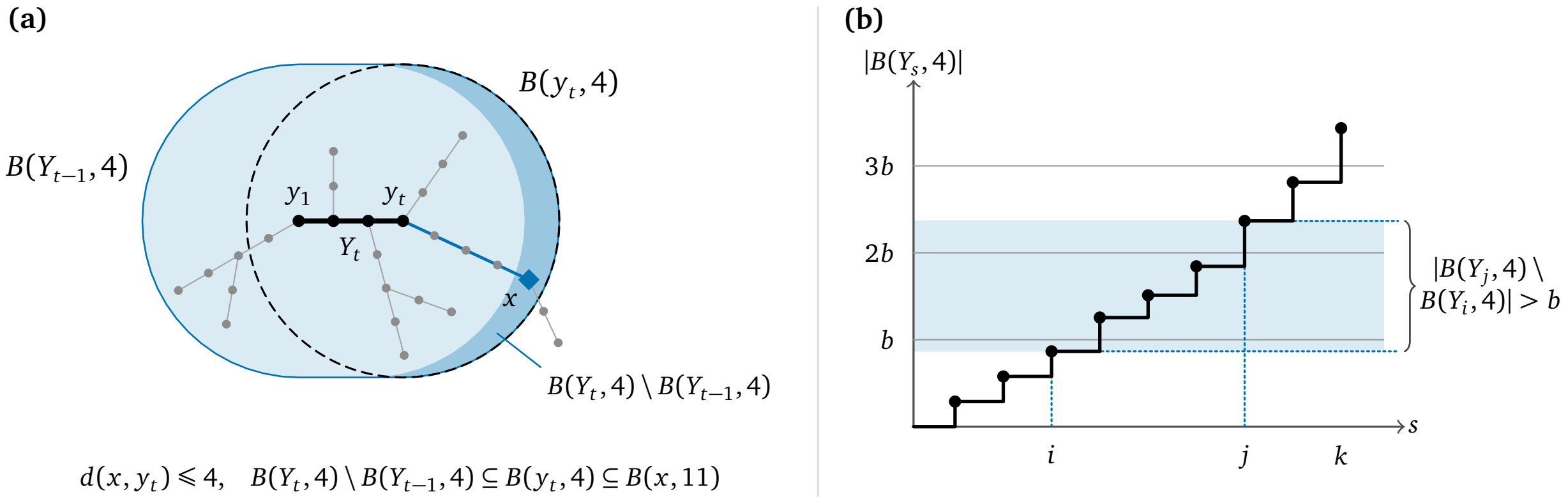

\begin{lemma}\label{lm:find-prefix} Let  $Y = \{y_1,\ldots, y_k\}$ be an (ordered) set of $k$ vertices such that $|B(Y,4)| > b$. Then there is a randomized algorithm, denoted by $\textsc{FindPrefix}(Y)$, that returns either (i) a vertex $x\in V$ such that $|B(x,11)| > b$ or (ii) an index $t\in [k]$ such that:
\begin{equation*}
    b < |B(\{y_1,\ldots, y_t\}, 4)| \leq 3b~.
\end{equation*}
The algorithm makes $O(h^2)$ single-source shortest-path computations and spends $O(h^2(m+n))$ additional time, in expectation.
\end{lemma}
\begin{proof}  For notational convenience, we also allow the index $t=0$. For every $t\in [0,k]$, let $Y_t = \{y_1,\ldots, y_t\}$ be the prefix of $Y$ with $t$ elements; $Y_k = Y$ and $Y_0 = \varnothing$. Note that $B(Y_0, 4) = \varnothing$.   \Cref{alg:find-prefix} implements $\textsc{FindPrefix}(Y)$; see \Cref{fig:find-prefix}.   The idea is to sample a vertex $x \in V$ and check if $|B(x,11)| > b$. If so, we have found $x$ as in the lemma and return it. Otherwise, let $t$ be the smallest index in $[k]$ such that $d(x,y_t) \leq 4$. (If no such $t$ exists, then \Cref{alg:find-prefix} would set $t = 0$, and the algorithm will try sampling $x$ again since in this case $|B(Y_t,4)| = |B(Y_0, 4)| =  0$.) In the next iteration, the algorithm will continue checking if $Y_t$ satisfies the condition of the lemma, and if so, return it.
 
\begin{algorithm}[H]
\caption{$\textsc{FindPrefix}(Y \coloneqq \{y_1,\ldots, y_k\})$}\label{alg:find-prefix}
\begin{algorithmic}[1]
      \State $t\leftarrow k$
        \While{$|B(Y_t, 4)|\leq b$ or $|B(Y_t,4)|>3b$}
            \State sample $x\in V$ uniformly at random \label{line:y-sample}
            \If{$|B(x,11)|>b$}\label{line:bop-y}
                \State \Return $x$
            \EndIf
             \State $t\leftarrow$ the smallest index in $[k]$ such that $d(x,y_t)\leq 4$, or $t\leftarrow 0$ if there is none \label{line:t-exists}
        \EndWhile
        \State return $t$
\end{algorithmic}
\end{algorithm}

Now we analyze \Cref{alg:find-prefix}. Let $i \in [0,k-1]$ be the \underline{largest index} such that $|B(Y_i, 4)|\leq b$.   Let $j \in [1,k]$ be the \underline{smallest index} such that $|B(Y_j, 4)| > 2b$. If $|B(Y,4)|\leq 3b$, the algorithm returns $t=k$ immediately, so assume $|B(Y,4)|>3b$. Then $i\neq k$, $j$ exists, and $i<j$.

Observe that $|B(Y_j, 4) \setminus B(Y_i, 4)| > b = n/(8h^2)$. Then with probability at least $1/(8h^2)$, we sample $x\in B(Y_j, 4) \setminus B(Y_i, 4)$ in line~\ref{line:y-sample}. In this case, $d(x,Y) \leq 4$ and hence $t$ in line~\ref{line:t-exists} exists. Furthermore, $i < t \leq j$ since $d(x,Y_i) > 4$ by the definition of $x$. Thus,  by the maximality of $i$, $|B(Y_t, 4)| > b$. 

It remains to show that $|B(Y_t, 4)| \leq  3b$ under the assumption that  $|B(x,11)| \leq b$ (otherwise the algorithm correctly returns $x$). By the minimality of $j$, $|B(Y_{t-1},4)| \leq 2b$. Observe that  since  $B(y_t,4) \subseteq B(x,11)$  and  $|B(x,11)|\leq b$, it holds that $|B(y_t,4)| \leq b$. Thus:
\begin{equation*}
    |B(Y_t,4)| \leq  |B(Y_{t-1},4)| + |B(y_t,4)| \leq 2b + b = 3b 
\end{equation*}
as claimed.

The running time follows from the fact that we have $O(h^2)$ iterations in expectation since, as noted above,  we (successfully) sample $x\in B(Y_j, 4) \setminus B(Y_i, 4)$ in line~\ref{line:y-sample} with probability at least $1/(8h^2)$. In each iteration, we only need two shortest path computations: one from $x$ and another to compute $B(Y_t,4)$ (by adding a dummy vertex and connecting it to every vertex of $Y_t$).
\end{proof}

Now we give the algorithm claimed in \Cref{lem:ball-or-pair}.  \Cref{alg:ball-or-pair} tries to build an induced minor model of $K_h^{(1)}$, and if it fails to do so, it will either output a vertex $x$ or a pair $(A,B)$ as in \Cref{lem:ball-or-pair}. The algorithm has two major steps.

\begin{algorithm}[!t]
\caption{$\textsc{BallOrPair}(G,h)$}\label{alg:ball-or-pair}
\begin{algorithmic}[1]
    \State $b:=n/(8h^2)$
    \For{$i=1,\ldots,h$} \Comment{\lightgray{Step 1}}
        \State $F\leftarrow X_1\cup\cdots\cup X_{i-1}$
        \State $Y\leftarrow$ largest connected component of $G\setminus B(F,3)$
        \If{$|V(Y)| \leq b$} \Comment{\lightgray{Case 1, Step 1}}
            \State $U\leftarrow \varnothing$
            \For{each component $K$ of   $G\setminus B(F,3)$}
               \State $U\leftarrow U\cup (V(K)\setminus B(F,4))$
                \If{$|U| > b$}
                    return $(U,V\setminus(B(F,4)\cup U))$\label{line:U-return}
                \EndIf
            \EndFor
        \EndIf

        \State $y_1,\ldots,y_k\leftarrow$ a depth-first search order of $Y$ with $k = |V(Y)|$ \Comment{\lightgray{Case 2, Step 1}}
        \State $(x,t)\leftarrow$ \textsc{FindPrefix}$(\{y_1,\ldots,y_k\})$ \Comment{\lightgray{In \Cref{lm:find-prefix}}}
        \If{$x\not=$ \textsc{Null}} 
             return $x$
        \EndIf
        \State $X_i:=\{y_1,\ldots,y_t\}$\label{line:bop-Xi}
        \If{$|V\setminus B(X_i,5)|>b$} \label{line:last-cond}
            \State \Return $(B(X_i,4), V\setminus B(X_i,5))$\label{line:bop-far} \Comment{\lightgray{End of Step 1}}
        \EndIf
    \EndFor  
    \State $C\leftarrow\{v\in V:4<d(v,X_i)\leq 5\text{ for every }i\in [h]\}$\label{line:bop-C} \Comment{\lightgray{Step 2}}
    \For{every pair $1\leq i<j\leq h$}
        \State $c_{ij}:=$ any vertex of $C$
        \If{$|B(c_{ij},11)|>b$}\label{line:bop-c}
            \State \Return $c_{ij}$
        \EndIf
        \State $C\leftarrow C\setminus B(c_{ij},11)$
        \State $P_{ij}\leftarrow$  the vertices of shortest paths from $c_{ij}$ to $X_i$ and to $X_j$, except their last vertices  \Comment{\lightgray{End of Step 2}}
    \EndFor
    \State \Return the induced minor model with branch sets $X_1,\ldots,X_h$ and $P_{ij}$ for $i<j$
\end{algorithmic}
\end{algorithm}

\begin{figure}[t]
\centering
\input{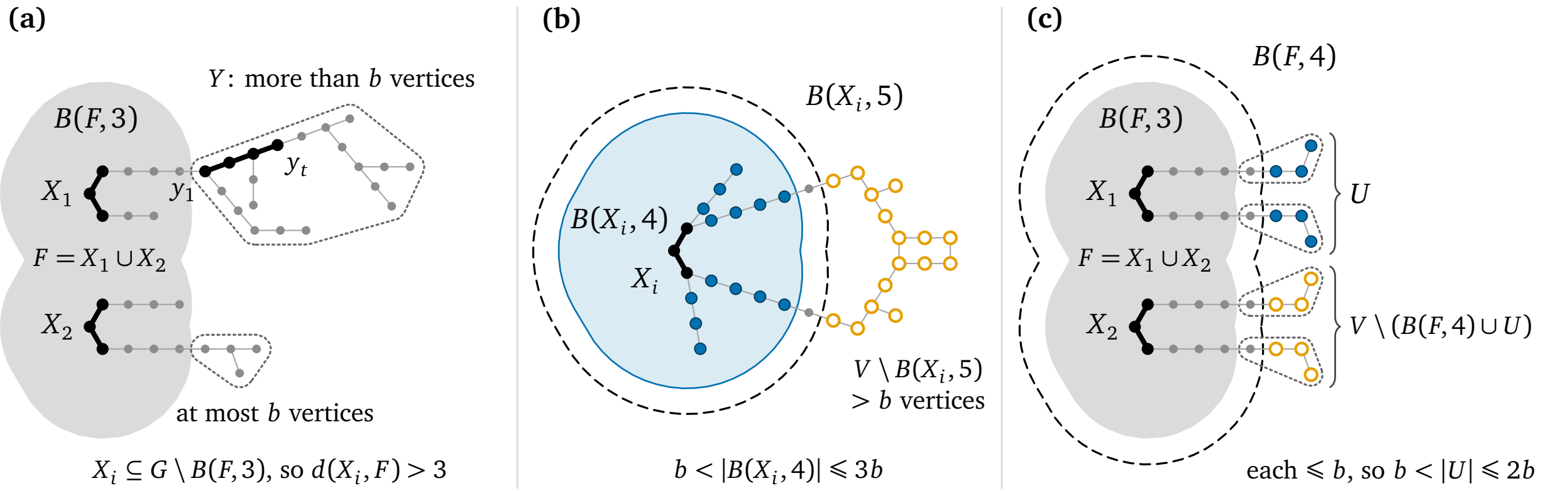}
\caption{Step 1 of \textsc{BallOrPair} (\Cref{alg:ball-or-pair}); all edges have length $1$. (a) Case 2: $F=X_1\cup X_2$ is the union of the branch sets found so far, the largest component $Y$ of $G\setminus B(F,3)$ has more than $b$ vertices, and $X_i=\{y_1,\ldots,y_t\}$ is a prefix of a depth-first search order of $Y$. (b) The check in line~\ref{line:last-cond}: if more than $b$ vertices (hollow) lie outside $B(X_i,5)$, the pair $(B(X_i,4),V\setminus B(X_i,5))$ is returned in line~\ref{line:bop-far}; the annulus between the two balls has width $1$, so the two sets are at distance more than $1$. (c) Case 1: every component of $G\setminus B(F,3)$ has at most $b$ vertices; their parts outside $B(F,4)$ are collected into $U$ (filled) until $|U|>b$, and the pair $(U,V\setminus(B(F,4)\cup U))$ is returned in line~\ref{line:U-return}.}
\label{fig:bop-growth}
\end{figure}

\paragraph{Step 1.} It finds a set of $h$ sets $X_1,\ldots, X_h$ (see \Cref{fig:bop-growth}) such that:
    \begin{equation}\label{eq:branch-sets}
    d(X_i,X_j)>3\quad(i\neq j),\qquad b<|B(X_i,4)|\leq 3b,\qquad|V\setminus B(X_i,5)|\leq b
\end{equation}
   Each set corresponds to a vertex of $K_h$ (but not a subdivision vertex of $K^{(1)}_h$). The idea is to find one set at a time: at iteration $i\in [h]$, we find a subset $X_i$ in $G\setminus B(F,3)$ where \EMPH{$F = X_1\cup\ldots\cup X_{i-1}$}. This is to guarantee that  $  d(X_i,X_j)>3\quad(i\neq j)$.  There are two cases.
\begin{itemize}
    \item \textbf{Case 1.~} If every connected component of $G\setminus B(F,3)$ has size at most $b$, then add the vertices of each connected component $K$ of $G\setminus B(F,3)$ \EMPH{outside $B(F,4)$} to $U$, one at a time until $|U| > b$. (Our analysis below shows that $U$ exists.) Then we return $(U,V\setminus(B(F,4)\cup U))$ as the pair. The reason that  $d(U,V\setminus(B(F,4)\cup U)) > 1$ is because every path of length at most $1$ that starts outside $B(F,4)$ stays outside $B(F,3)$.
    \item  \textbf{Case 2.~} The largest connected component $Y$ of   $G\setminus B(F,3)$ has more than $b$ vertices. Then we form a set $\{y_1,\ldots, y_k\}$ of its vertices ordered by DFS so that every prefix induces a connected subgraph of $Y$. Then we apply \Cref{lm:find-prefix} to find either (i) a vertex $x$ such that  $|B(x,11)| > b$ or (ii) a prefix $X_i = \{y_1,\ldots, y_t\}$ such that $ b<|B(X_i,4)|\leq 3b$. In the former case, we return $x$ as it satisfies \Cref{lem:ball-or-pair}. In the latter case, we have found a set $X_i$ that satisfies the second condition of \Cref{eq:branch-sets}. The only missing condition is $|V\setminus B(X_i,5)|\leq b$, and it is guaranteed by line~\ref{line:last-cond}. Specifically, if  $|V\setminus B(X_i,5)|> b$, then we can return $(B(X_i,4), V\setminus B(X_i,5))$ as the pair in \Cref{lem:ball-or-pair} since the two sets are at distance greater than $1$, and each of size at least $b$.
\end{itemize}
   
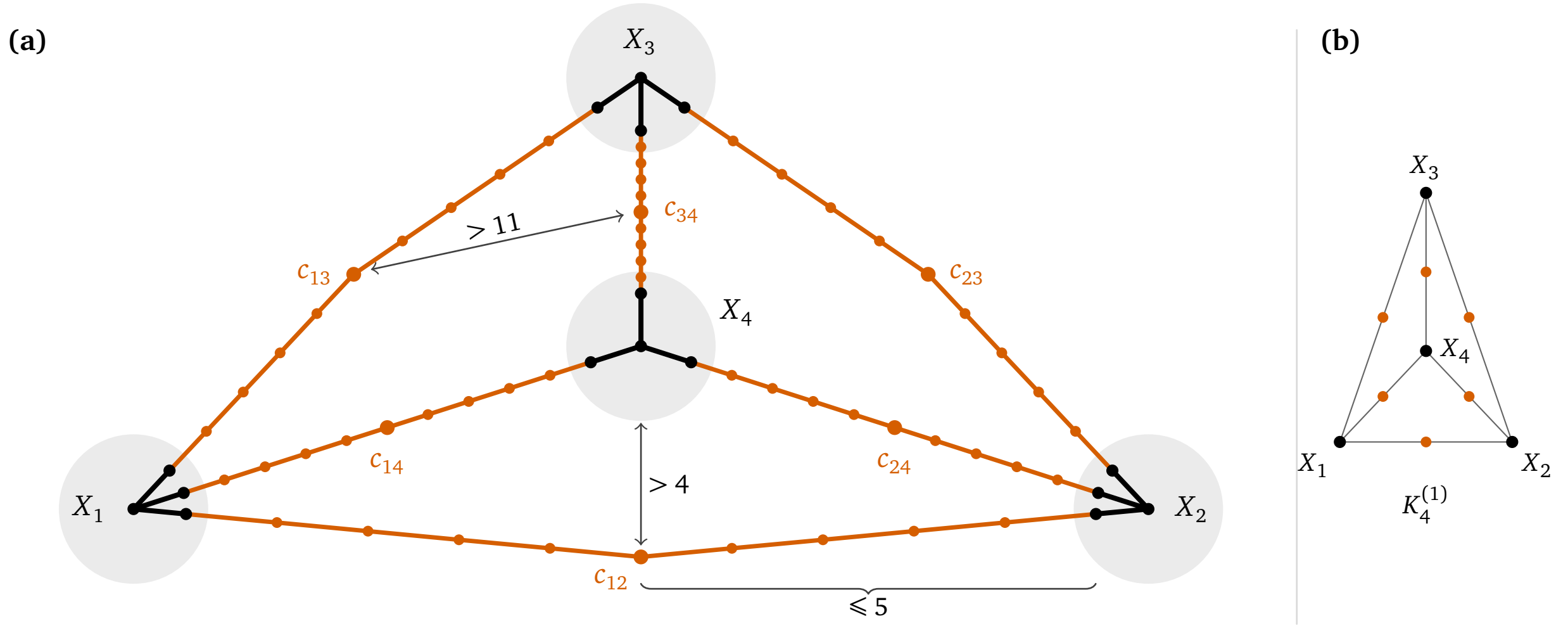
\begin{figure}[t]
\centering
% Step 2 of BallOrPair for h=4: branch sets X_1..X_4 (black claws on grey discs),
% centers c_ij (vermilion) and the sets P_ij: two paths of five unit edges from c_ij to X_i and X_j.
% Euclidean lengths are schematic.
\begin{tikzpicture}[figure]
\path[use as bounding box] (0,0.2) rectangle (16.40,6.80);
\draw[divider] (13.55,0.30)--(13.55,6.50);
\node[panel,anchor=west] at (0.04,6.38) {\textbf{(a)}};
\node[panel,anchor=west] at (13.75,6.38) {\textbf{(b)}};
\fill[branch] (1.4,1.5) circle (0.78);
\fill[branch] (12.0,1.5) circle (0.78);
\fill[branch] (6.7,6.0) circle (0.78);
\fill[branch] (6.7,3.2) circle (0.78);
\draw[measure] (6.7,1.12)--(6.7,2.40) node[midway,right=1pt,dim] {$>4$};
\draw[measure] (3.88,4.00)--(6.52,4.57) node[midway,sloped,above=1pt,dim] {$>11$};
\draw[brace] (11.45,0.72)--(6.70,0.72) node[midway,below=3pt,dim] {$\le 5$};
\draw[conn] (6.700,1.000)--(5.750,1.090)--(4.799,1.179)--(3.849,1.269)--(2.898,1.359)--(1.948,1.448);
\node[svtx] at (5.750,1.090) {};
\node[svtx] at (4.799,1.179) {};
\node[svtx] at (3.849,1.269) {};
\node[svtx] at (2.898,1.359) {};
\draw[conn] (6.700,1.000)--(7.650,1.090)--(8.601,1.179)--(9.551,1.269)--(10.502,1.359)--(11.452,1.448);
\node[svtx] at (7.650,1.090) {};
\node[svtx] at (8.601,1.179) {};
\node[svtx] at (9.551,1.269) {};
\node[svtx] at (10.502,1.359) {};
\draw[conn] (3.700,3.950)--(3.315,3.540)--(2.931,3.130)--(2.546,2.721)--(2.161,2.311)--(1.776,1.901);
\node[svtx] at (3.315,3.540) {};
\node[svtx] at (2.931,3.130) {};
\node[svtx] at (2.546,2.721) {};
\node[svtx] at (2.161,2.311) {};
\draw[conn] (3.700,3.950)--(4.209,4.298)--(4.718,4.646)--(5.228,4.994)--(5.737,5.342)--(6.246,5.690);
\node[svtx] at (4.209,4.298) {};
\node[svtx] at (4.718,4.646) {};
\node[svtx] at (5.228,4.994) {};
\node[svtx] at (5.737,5.342) {};
\draw[conn] (9.700,3.950)--(10.085,3.540)--(10.469,3.130)--(10.854,2.721)--(11.239,2.311)--(11.624,1.901);
\node[svtx] at (10.085,3.540) {};
\node[svtx] at (10.469,3.130) {};
\node[svtx] at (10.854,2.721) {};
\node[svtx] at (11.239,2.311) {};
\draw[conn] (9.700,3.950)--(9.191,4.298)--(8.682,4.646)--(8.172,4.994)--(7.663,5.342)--(7.154,5.690);
\node[svtx] at (9.191,4.298) {};
\node[svtx] at (8.682,4.646) {};
\node[svtx] at (8.172,4.994) {};
\node[svtx] at (7.663,5.342) {};
\draw[conn] (4.050,2.350)--(3.625,2.214)--(3.199,2.077)--(2.774,1.941)--(2.349,1.804)--(1.924,1.668);
\node[svtx] at (3.625,2.214) {};
\node[svtx] at (3.199,2.077) {};
\node[svtx] at (2.774,1.941) {};
\node[svtx] at (2.349,1.804) {};
\draw[conn] (4.050,2.350)--(4.475,2.486)--(4.901,2.623)--(5.326,2.759)--(5.751,2.896)--(6.176,3.032);
\node[svtx] at (4.475,2.486) {};
\node[svtx] at (4.901,2.623) {};
\node[svtx] at (5.326,2.759) {};
\node[svtx] at (5.751,2.896) {};
\draw[conn] (9.350,2.350)--(9.775,2.214)--(10.201,2.077)--(10.626,1.941)--(11.051,1.804)--(11.476,1.668);
\node[svtx] at (9.775,2.214) {};
\node[svtx] at (10.201,2.077) {};
\node[svtx] at (10.626,1.941) {};
\node[svtx] at (11.051,1.804) {};
\draw[conn] (9.350,2.350)--(8.925,2.486)--(8.499,2.623)--(8.074,2.759)--(7.649,2.896)--(7.224,3.032);
\node[svtx] at (8.925,2.486) {};
\node[svtx] at (8.499,2.623) {};
\node[svtx] at (8.074,2.759) {};
\node[svtx] at (7.649,2.896) {};
\draw[conn] (6.700,4.600)--(6.700,4.770)--(6.700,4.940)--(6.700,5.110)--(6.700,5.280)--(6.700,5.450);
\node[svtx] at (6.700,4.770) {};
\node[svtx] at (6.700,4.940) {};
\node[svtx] at (6.700,5.110) {};
\node[svtx] at (6.700,5.280) {};
\draw[conn] (6.700,4.600)--(6.700,4.430)--(6.700,4.260)--(6.700,4.090)--(6.700,3.920)--(6.700,3.750);
\node[svtx] at (6.700,4.430) {};
\node[svtx] at (6.700,4.260) {};
\node[svtx] at (6.700,4.090) {};
\node[svtx] at (6.700,3.920) {};
\draw[bed] (1.400,1.500)--(1.948,1.448); \node[bvtx] at (1.948,1.448) {};
\draw[bed] (1.400,1.500)--(1.776,1.901); \node[bvtx] at (1.776,1.901) {};
\draw[bed] (1.400,1.500)--(1.924,1.668); \node[bvtx] at (1.924,1.668) {};
\node[bvtx] at (1.400,1.500) {};
\draw[bed] (12.000,1.500)--(11.452,1.448); \node[bvtx] at (11.452,1.448) {};
\draw[bed] (12.000,1.500)--(11.624,1.901); \node[bvtx] at (11.624,1.901) {};
\draw[bed] (12.000,1.500)--(11.476,1.668); \node[bvtx] at (11.476,1.668) {};
\node[bvtx] at (12.000,1.500) {};
\draw[bed] (6.700,6.000)--(6.246,5.690); \node[bvtx] at (6.246,5.690) {};
\draw[bed] (6.700,6.000)--(7.154,5.690); \node[bvtx] at (7.154,5.690) {};
\draw[bed] (6.700,6.000)--(6.700,5.450); \node[bvtx] at (6.700,5.450) {};
\node[bvtx] at (6.700,6.000) {};
\draw[bed] (6.700,3.200)--(6.176,3.032); \node[bvtx] at (6.176,3.032) {};
\draw[bed] (6.700,3.200)--(7.224,3.032); \node[bvtx] at (7.224,3.032) {};
\draw[bed] (6.700,3.200)--(6.700,3.750); \node[bvtx] at (6.700,3.750) {};
\node[bvtx] at (6.700,3.200) {};
\node[ctr] at (6.700,1.000) {};
\node[ctr] at (3.700,3.950) {};
\node[ctr] at (9.700,3.950) {};
\node[ctr] at (4.050,2.350) {};
\node[ctr] at (9.350,2.350) {};
\node[ctr] at (6.700,4.600) {};
\node[lab,anchor=east] at (1.15,1.5) {$X_1$};
\node[lab,anchor=west] at (12.25,1.5) {$X_2$};
\node[lab,anchor=south] at (6.7,6.2) {$X_3$};
\node[lab,anchor=west] at (7.5,3.55) {$X_4$};
\node[path text,anchor=north east] at (6.62,0.92) {$c_{12}$};
\node[path text,anchor=east] at (3.52,3.95) {$c_{13}$};
\node[path text,anchor=west] at (9.88,3.95) {$c_{23}$};
\node[path text,anchor=north] at (4.05,2.13) {$c_{14}$};
\node[path text,anchor=north] at (9.35,2.13) {$c_{24}$};
\node[path text,anchor=west] at (6.9,4.6) {$c_{34}$};
\begin{scope}[xshift=-0.25cm]
\draw[ed,black!60] (14.250,2.200)--(16.050,2.200); \node[svtx] at (15.150,2.200) {};
\draw[ed,black!60] (14.250,2.200)--(15.150,4.800); \node[svtx] at (14.700,3.500) {};
\draw[ed,black!60] (16.050,2.200)--(15.150,4.800); \node[svtx] at (15.600,3.500) {};
\draw[ed,black!60] (14.250,2.200)--(15.150,3.150); \node[svtx] at (14.700,2.675) {};
\draw[ed,black!60] (16.050,2.200)--(15.150,3.150); \node[svtx] at (15.600,2.675) {};
\draw[ed,black!60] (15.150,4.800)--(15.150,3.150); \node[svtx] at (15.150,3.975) {};
\node[bvtx] at (14.250,2.200) {};
\node[bvtx] at (16.050,2.200) {};
\node[bvtx] at (15.150,4.800) {};
\node[bvtx] at (15.150,3.150) {};
\node[dim,anchor=north east] at (14.18,2.12) {$X_1$};
\node[dim,anchor=north west] at (16.12,2.12) {$X_2$};
\node[dim,anchor=south] at (15.15,4.9) {$X_3$};
\node[dim,anchor=west] at (15.27,3.15) {$X_4$};
\node[dim] at (15.15,1.55) {$K_4^{(1)}$};
\end{scope}
\end{tikzpicture}%
\caption{Step 2 of \textsc{BallOrPair} and the resulting induced minor model of $K^{(1)}_h$, drawn for $h=4$. (a) The branch sets $X_1,\ldots,X_4$ (black) are connected and pairwise at distance more than $3$. Each center $c_{ij}$ (orange) is at distance in $(4,5]$ from every $X_\ell$, and the centers are pairwise at distance more than $11$. The set $P_{ij}$ (orange) consists of $c_{ij}$ and the shortest paths from $c_{ij}$ to $X_i$ and to $X_j$ without their last vertices, so it is adjacent to $X_i$ and $X_j$ and to no other branch set. (b) Contracting each $X_i$ to a vertex and each $P_{ij}$ to a vertex gives $K^{(1)}_4$.}
\label{fig:induced-minor-model}
\end{figure}

\paragraph{Step 2.} This step finds a set $\{c_{ij}\}_{1\leq i< j \leq h}$ (see \Cref{fig:induced-minor-model}) corresponding to the subdivision vertices of $K^{(1)}_h$ in the induced minor model, as well as a vertex subset $P_{ij}$ of the two shortest paths from $c_{ij}$ to $X_i$ and $X_j$ (without their last vertices in  $X_i$ and $X_j$). Each $c_{ij}$ belongs to the set $C = \{v\in V:4<d(v,X_i)\leq 5\text{ for every }i\in [h]\}$ and hence:
\begin{equation}\label{eq:dist-cij-X}
    4<d(c_{ij},X_\ell) \leq 5  \text{ for every } \ell \in [h]
\end{equation}

To get an induced minor model of $K^{(1)}_h$, we want $\{P_{ij}\}_{1\leq i < j\leq h}$ to be non-adjacent. We enforce this indirectly by making sure that:
\begin{equation}\label{eq:dist-cij-cij}
    d(c_{ij}, c_{\Tilde{i}\Tilde{j}}) > 11 \text{ for } \{i,j\}\not= \{\Tilde{i}, \Tilde{j}\}.
\end{equation}
Specifically, every time we take a new vertex $c_{ij}$ from $C$, we update $C\leftarrow C\setminus B(c_{ij}, 11)$. If $ |B(c_{ij}, 11)| > b$, then we return $c_{ij}$ as the vertex in \Cref{lem:ball-or-pair}; otherwise, we can show that $C\not=\varnothing$ after the update, and hence we can continue the next iterations. 

As long as $G$ is $K^{(1)}_h$-induced-minor-free, we will never reach the last line of \Cref{alg:ball-or-pair} and hence the algorithm always either returns a vertex $x$ or a pair $(A,B)$ as required by \Cref{lem:ball-or-pair}. 

\begin{proofof}{\Cref{lem:ball-or-pair}} First, observe that \Cref{eq:branch-sets} is satisfied for $\{X_1,\ldots,X_h\}$ (if the algorithm does not terminate early): $d(X_i,X_k) > 3$ since $X_i$ is in $G\setminus B(F,3)$, $b< |B(X_i,4)|\leq 3b$ by \Cref{lm:find-prefix}, and $|V\setminus B(X_i,5)|\leq b$ due to line~\ref{line:last-cond}.

We have argued in Case 1 that $d(U,V\setminus(B(F,4)\cup U)) > 1$. To show that this pair satisfies \Cref{lem:ball-or-pair}, it remains to show that each set has size at least $b$. By construction, $b< |U| \leq 2b$ since every connected component of $G\setminus B(F,3)$ has at most $b$ vertices. Since $|V\setminus B(F,4)|\geq n- h(3b)>3b$ by \eqref{eq:branch-sets}, more than $b$ vertices of $V\setminus B(F,4)$ are not in $U$. Thus, the pair returned in line~\ref{line:U-return} is correct. If the algorithm returns a pair  $(B(X_i,4), V\setminus B(X_i,5))$ in line~\ref{line:bop-far}, we have argued in Case 2 that the two sets are at distance greater than $1$, and each of size at least $b$.

If the algorithm returns $c_{ij}$ in line~\ref{line:bop-c}, then it is correct as $|B(c_{ij},11)|>b$. Thus, for correctness, we only need to argue that if no $c_{ij}$  is returned, Step 2 would output the induced minor model of $K^{(1)}_h$ (which would contradict that $G$ is $K^{(1)}_h$-induced-minor-free). 

First, we show that the algorithm runs exactly $\binom{h}{2}$ iterations in Step 2 by showing that $C$ is never empty. We note that $B(X_i,5)$ contains at least $n-b$ vertices and since $|B(X_i,4)|\leq 3b$, $B(X_i,5)\setminus B(X_i,4)$ misses at most $b + 3b = 4b$ vertices of $V$. Thus, over $h$ sets, $|C|\geq n-h(4b)\geq 3n/4$. In each iteration, we remove at most $b$ vertices from $C$, over $\binom{h}{2}$ iterations, we remove:
\begin{equation*}
    \binom{h}{2}b =   \binom{h}{2} \frac{n}{8h^2}<n/16
\end{equation*}
vertices from $C$. This means $C$ will never be empty.

It remains to show that $\{X_i\}_{1\leq i\leq h}\cup \{P_{ij}\}_{1\leq i < j \leq h}$ form an induced minor model of $K^{(1)}_h$ in $G$. Recall that the edge lengths are in $(0,1]$.  For every $1\leq i < j \leq h$, $\ell \in [h]\setminus \{i,j\}$ and every neighbor $u$ of $X_{\ell}$ in $G$:

\begin{equation*}
   \begin{split}
        d(c_{ij},u) &\geq d(c_{ij},X_\ell)- d(u,X_{\ell}) > 4-1 = 3  \\
        d(u,X_i) &\geq d(X_\ell,X_i)-1>3-1 = 2\\
        d(u,X_j) &\geq d(X_\ell,X_j)-1>3-1 = 2
   \end{split}
\end{equation*}
This means $P_{ij}$ cannot contain $u$ since $d(c_{ij},X_i), d(c_{ij},X_j) \leq 5$ by \Cref{eq:dist-cij-X}. Thus, $P_{ij}$ is disjoint from $X_i$ and $X_j$, has neighbors in both, and is not adjacent to any other branch sets.  

Distinct sets $P_{ij} \subseteq B(c_{ij}, 5)$ and by \Cref{eq:dist-cij-cij} and the triangle inequality, $P_{ij}$ and $P_{\Tilde{i}\Tilde{j}}$ are at distance more than $1$ and hence they are non-adjacent. Thus,  $\{X_i\}_{1\leq i\leq h}\cup \{P_{ij}\}_{1\leq i < j \leq h}$ form an induced minor model of $K^{(1)}_h$ in $G$, as claimed.

By \Cref{lm:find-prefix}, finding each $X_i$ takes $O(h^2)$ shortest-path computations in expectation.  Step 2 also requires only $O(h^2)$ shortest-path computations: one per $c_{ij}$. Thus, the overall number of shortest-path computations is $O(h^3)$.
\end{proofof}

\bibliographystyle{alphaurl}
\bibliography{ref}

\end{document}